\documentclass[12pt,preprint]{aastex62}

\newcommand{\be}{\begin{eqnarray}}
\newcommand{\ee}{\end{eqnarray}}

\newcommand{\revise}[1]{#1}
\newcommand{\ISOIS}{IS$\odot$IS}

\begin{document} 

\title{Seed Population Pre-Conditioning and Acceleration Observed by
  Parker Solar Probe}

\author{N. A. Schwadron}
\affiliation{University of New Hampshire, Durham, NH, 03824, USA}
\affiliation{Department of Astrophysical Sciences, Princeton University, Princeton, NJ, 08544, USA}

\author{S. Bale}
\affiliation{University of California at Berkeley, Berkeley, CA 94720, USA}

\author{J. Bonnell}
\affiliation{University of California at Berkeley, Berkeley, CA 94720, USA}

\author{A. Case}
\affiliation{Harvard-Smithsonian Center for Astrophysics, Cambridge, MA 02138, USA}

\author{E. R. Christian}
\affiliation{Goddard Space Flight Center, Greenbelt, MD 20771, USA}

\author{C. M. S. Cohen}
\affiliation{California Institute of Technology, Pasadena, CA 91125, USA}

\author{A. C.  Cummings}
\affiliation{California Institute of Technology, Pasadena, CA 91125, USA}

\author{A. J.  Davis}
\affiliation{California Institute of Technology, Pasadena, CA 91125, USA}

\author{T. Dudok de Wit}
\affiliation{LPC2E, CNRS and University of Orl\'eans, Orl\'eans, France}

\author{W. de Wet}
\affiliation{University of New Hampshire, Durham, NH, 03824, USA}

\author{M. I. Desai}
\affiliation{University of Texas at San Antonio, San Antonio, TX 78249, USA }

\author{C. J. Joyce}
\affiliation{Department of Astrophysical Sciences, Princeton University, Princeton, NJ, 08544, USA}

\author{K. Goetz}
\affiliation{University of California at Berkeley, Berkeley, CA 94720, USA}

\author{J. Giacalone}
\affiliation{University of Arizona, Tucson, AZ 85721, USA }

\author{M. Gorby}
\affiliation{University of New Hampshire, Durham, NH, 03824, USA}

\author{P. Harvey}
\affiliation{University of California at Berkeley, Berkeley, CA 94720, USA}

\author{B. Heber}
 \affiliation{Christian-Albrechts-University of Kiel, Kiel, 24118, Germany} 

\author{M. E. Hill}
\affiliation{Applied Physics Laboratory, Laurel, MD 20723, USA}

\author{M. Karavolos}
\affiliation{Institute for Astronomy,
  Astrophysics, Space Applications and Remote Sensing of the National
  Observatory of Athens, Vas. Pavlou and I. Metaxa, 15236 Penteli,
  Greece}

\author{J. C. Kasper}
\affiliation{University of Michigan, Ann Arbor, MI 48109, USA}

\author{K. Korreck}
\affiliation{Harvard-Smithsonian Center for Astrophysics, Cambridge, MA 02138, USA}

\author{D. Larson}
\affiliation{University of California at Berkeley, Berkeley, CA 94720, USA}

\author{R. Livi}
\affiliation{University of California at Berkeley, Berkeley, CA 94720, USA}

\author{R. A. Leske}
\affiliation{California Institute of Technology, Pasadena, CA 91125, USA}

\author{O. Malandraki} 
\affiliation{Institute for Astronomy,
  Astrophysics, Space Applications and Remote Sensing of the National
  Observatory of Athens, Vas. Pavlou and I. Metaxa, 15236 Penteli,
  Greece}

\author{R. MacDowall}
\affiliation{Goddard Space Flight Center, Greenbelt, MD 20771, USA}

\author{D. Malaspina} 
\affiliation{Laboratory for Atmospheric and
  Space Physics, University of Colorado, Boulder, CO 80303, USA}

\author{W. H. Matthaeus}
\affiliation{University of Delaware, Newark, DE 19716, USA}

\author{D. J. McComas} 
\affiliation{Department of Astrophysical Sciences, Princeton University, Princeton, NJ, 08544, USA}

\author{R. L. McNutt Jr.}
\affiliation{Applied Physics Laboratory, Laurel, MD 20723, USA}

\author{R. A. Mewaldt}
\affiliation{California Institute of Technology, Pasadena, CA 91125, USA}

\author{D. G. Mitchell}
\affiliation{Applied Physics Laboratory, Laurel, MD 20723, USA}

\author{L. Mays}
\affiliation{Goddard Space Flight Center, Greenbelt, MD 20771, USA}

\author{J. T. Niehof}
\affiliation{University of New Hampshire, Durham, NH, 03824, USA}

\author{D. Odstrcil}
\affiliation{Goddard Space Flight Center, Greenbelt, MD 20771, USA}

\author{M. Pulupa}
\affiliation{University of California at Berkeley, Berkeley, CA 94720, USA}

 \author{B. Poduval}
\affiliation{University of New Hampshire, Durham, NH, 03824, USA}

\author{J. S. Rankin}
\affiliation{Department of Astrophysical Sciences, Princeton University, Princeton, NJ, 08544, USA}

\author{E. C. Roelof}
\affiliation{Applied Physics Laboratory, Laurel, MD 20723, USA}

\author{M. Stevens}
\affiliation{Harvard-Smithsonian Center for Astrophysics, Cambridge, MA 02138, USA}

\author{E. C. Stone}
\affiliation{California Institute of Technology, Pasadena, CA, 91125, USA}

\author{J. R. Szalay}
\affiliation{Department of Astrophysical Sciences, Princeton University, Princeton, NJ, 08544, USA}

\author{M. E. Wiedenbeck}
\affiliation{California Institute of Technology, Pasadena, CA 91125, USA}

\author{R. Winslow}
\affiliation{University of New Hampshire, Durham, NH, 03824, USA}

\author{P. Whittlesey}
\affiliation{University of California at Berkeley, Berkeley, CA 94720, USA}



\begin{abstract} 
A series of solar energetic particle (SEP) events were observed at
Parker Solar Probe (PSP) by the Integrated Science Investigation of
the Sun (\ISOIS) during the period from April 18, 2019 through April
24, 2019.  The PSP spacecraft was located near 0.48 au from the Sun on
Parker spiral field lines that projected out to 1 au within $\sim
25^\circ$ of near Earth spacecraft. These SEP events, though small
compared to historically large SEP events, were amongst the largest
observed thus far in the PSP mission and provide critical information
about the space environment inside 1 au during SEP events.  During
this period the Sun released multiple coronal mass ejections
(CMEs). One of these CMEs observed was initiated on April 20, 2019 at
01:25 UTC, and the interplanetary CME (ICME) propagated out and passed
over the PSP spacecraft.  Observations by the Electromagnetic Fields
Investigation (FIELDS) show that the magnetic field structure was
mostly radial throughout the passage of the compression region and the
plasma that followed, indicating that PSP did not directly observe a
flux rope internal to the ICME, consistent with the location of PSP on
the ICME flank.  Analysis using relativistic electrons observed near
Earth by the Electron, Proton and Alpha Monitor (EPAM) on the Advanced
Composition Explorer (ACE) demonstrates the presence of electron seed
populations (40--300 keV) during the events observed.  The energy
spectrum of the \ISOIS~ observed proton seed population below 1 MeV is
close to the limit of possible stationary state plasma distributions
out of equilibrium.  \ISOIS~ observations reveal the
\revise{enhancement} of seed populations during the passage of the
ICME, which \revise{likely indicates a key part} of the
pre-acceleration process that occurs close to the Sun.
\end{abstract}


\section{Introduction}
The NASA Parker Solar Probe (PSP) Mission provides humanity's
first direct exploration of our Star, and its environment
\cite[]{Fox:2016}.  The Integrated Science Investigation of the Sun
(\ISOIS) instrument suite \cite[]{McComas:2016} provides
comprehensive measurements of solar energetic particles (SEPs) using
two Energetic Particle Instruments measuring higher (EPI-Hi) and
lower (EPI-Lo) energy particles \cite[]{McComas:2016} over the
range 0.02 -- 200 MeV/nucleon. Here, we examine the sources of
this energetic particle environment and the seed populations therein,
which respond dynamically to the solar wind observed by the Solar Wind
Electrons Alphas and Protons Investigation (SWEAP)
\cite[]{Kasper:2016} and the magnetic field observed by the
Electromagnetic Fields Investigation (FIELDS) \cite[]{Bale:2016}
locally around PSP. Global context for the solar wind's surrounding
density structures are observed by the Wide Field Imager for Solar
Probe Plus (WISPR) \cite[]{Vourlidas:2016}, though WISPR observations
not reported within this paper.

We use solar energetic particle events (SEPs) observed by \ISOIS~ to
examine the period from April 18, 2019 to April 24, 2019 when two
active regions near the Sun's equator became highly active,
releasing numerous flares and coronal mass ejections (CMEs). We
describe the SEP events observed by \ISOIS~ that occurred over this
period. These events are the largest SEP events so far
observed by PSP \cite[]{McComas:2019} and demonstrate the complex
interplay between flares, seed populations and coronal mass ejections
in the acceleration of energetic particles near the Sun.

The paper is organized as follows. In \S\ref{sec:analysis}, we provide
detailed analysis of the energetic particle events observed by \ISOIS~
from April 18 to 24, 2019. We then assess seed populations and
accelerated particles observed at PSP based on measurements of nearly
scatter-free electrons in \S\ref{sec:electrons}. Type III radio bursts
are associated with CMEs observed over this period.  We show results
of a CME model in \S\ref{sec:modeling}, and characterize the
compression of energetic particle seed populations during the passage
of an interplanetary CME over PSP.  In \S\ref{sec:conclusion}, we
summarize key results and present our major conclusions.

\section{SEPs observed by \ISOIS~ from April 18 to April 24, 2019}
\label{sec:analysis}

\begin{figure}
\includegraphics[width=0.95\columnwidth]{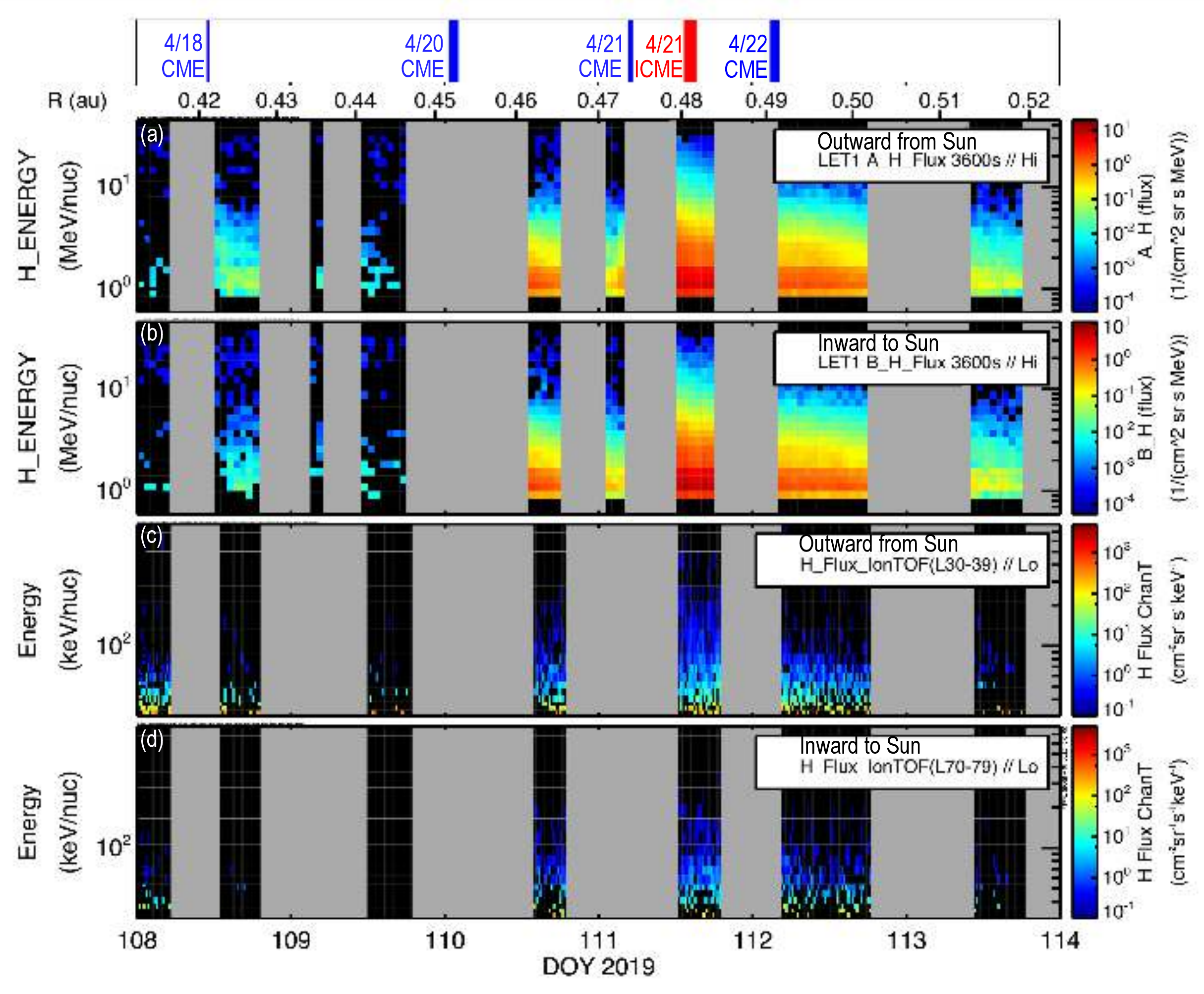} 
\caption{The energetic particle fluxes observed by \ISOIS~ during the
  period from April 18 to April 24, 2019. Panels (a) and (b) show
  energetic proton differential fluxes observed by EPI-Hi directed
  from and to the Sun on the nominal direction of Parker field
  lines. Panels (c) and (d) show EPI-Lo differential proton fluxes in
  the corresponding directions. The two top labels show radial
  distance and the start times of CME events observed by STEREO and
  LASCO. The CME start times and related information are listed in
  Table \ref{tab:table}. The width of the vertical lines are used to
  indicate the angular proximity of the CME to PSP, with the thickest
  lines corresponding the CMEs that likely overtook the PSP
  spacecraft.  Both CMEs released on April 20 and April 22 were
  directed such that they would overtake the PSP spacecraft.  The CME
  released on 4/20/2019 at 01:25 UTC drove a compression that overtook
  the PSP spacecraft on 4/21/2019 at $\sim$16:00 (red thick line). }
\label{fig:flux}
\end{figure}

The energetic particle fluxes observed from April 18 (day-of-year, DOY
108) to April 24 (DOY 114), 2019 are shown in Figure
\ref{fig:flux}. The energetic fluxes in the direction outward from the
Sun along a nominal Parker spiral (for a 400 km s$^{-1}$ solar wind
speed) are shown in Panel (a) for EPI-Hi and (c) for EPI-Lo.  Inward
fluxes along the nominal Parker spiral are shown in Panel (b) for
EPI-Hi and (d) for EPI-Lo.  Observed distributions early in the event
development (from DOY 108 through 109.8) show larger
anisotropies. These distributions become increasingly isotropic as the
associated SEP events progress after DOY 110.5.

PSP instruments were operational only intermittently during the period
studied. Satellite contacts including high-speed data transfers
occurred throughout the period. Instruments were powered off during
these periods.

\begin{figure}
\includegraphics[width=0.8\columnwidth]{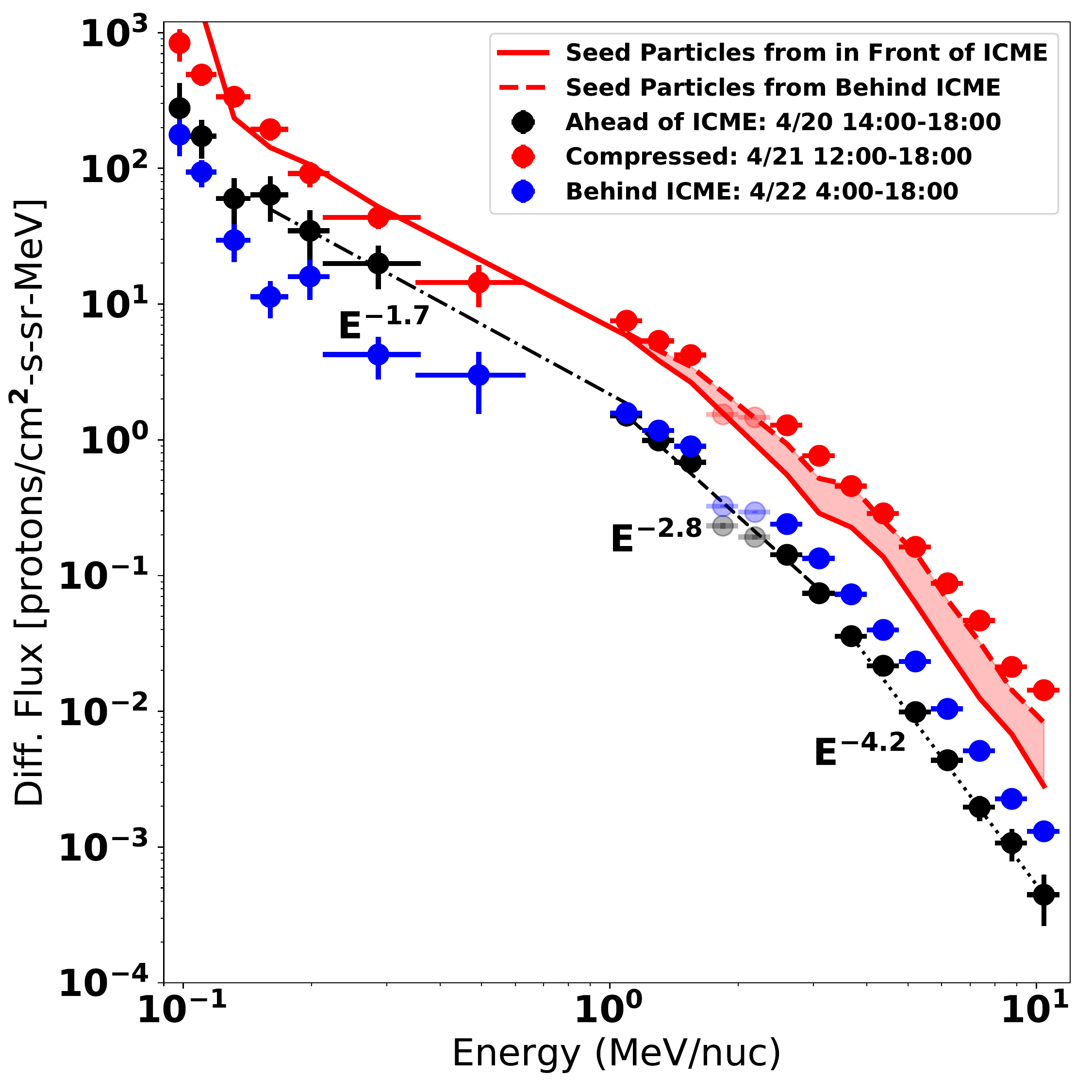}
\caption{The observed \ISOIS~ differential energy fluxes observed by
  before (black), during (red), and after the ICME passage (blue)
  plotted together with predictions (red curves) for compression of
  the energetic particle seed population based on detailed in \S
  \ref{sec:modeling}.  The semi-transparent data points below 2 MeV
  are likely affected by instrumental effects currently under
  study. Differential flux power-laws are shown by dot-dashed, dashed
  and dotted curves.  The solid and dashed red curves show prediction
  for compressed and accelerated seed populations using the energetic
  particle fluxes ahead and behind the ICME, respectively, as the
  proxy for the uncompressed seed population.  }
\label{fig:spectra2}
\end{figure}

The ion spectra averaged before (4/20 14:00 -- 18:00) during (4/21
12:00 -- 18:00) and after (4/22 04:00 -- 18:00) are shown in Figure
\ref{fig:spectra2}. We note that the spectra are all very similar, but
the differential energy spectrum observed during the ICME passage is
almost uniformly enhanced relative to the differential spectra
observed before and after the ICME passage. In \S\ref{sec:modeling},
we discuss modeling of compression and acceleration of the seed
populations swept up into the compression driven by an
ICME observed by PSP on April 21, 2019.

Four coronal mass ejections (CMEs) were released from the Sun on April
18, 20, 21, and 22.  These events were identified using the Space
Weather Database Of Notifications, Knowledge, Information (DONKI)
provided by the Community Coordinated Modeling Center (CCMC,
https://ccmc.gsfc.nasa.gov).  Table \ref{tab:table} provides details
on each of these events including the start time, direction, PSP
location, initial speed, and width of the CME.  The vertical lines in
the top panel of Figure \ref{fig:flux} identify these CME release
times, and the width of the vertical lines are used to indicate the
angular proximity of the CME to PSP, with the thickest lines
corresponding to CMEs that overtook the PSP spacecraft. In particular,
both CMEs released on April 20 and April 22, respectively, propagated
in a direction such that they would overtake the PSP spacecraft.

\begin{figure}
\includegraphics[width=0.95\columnwidth]{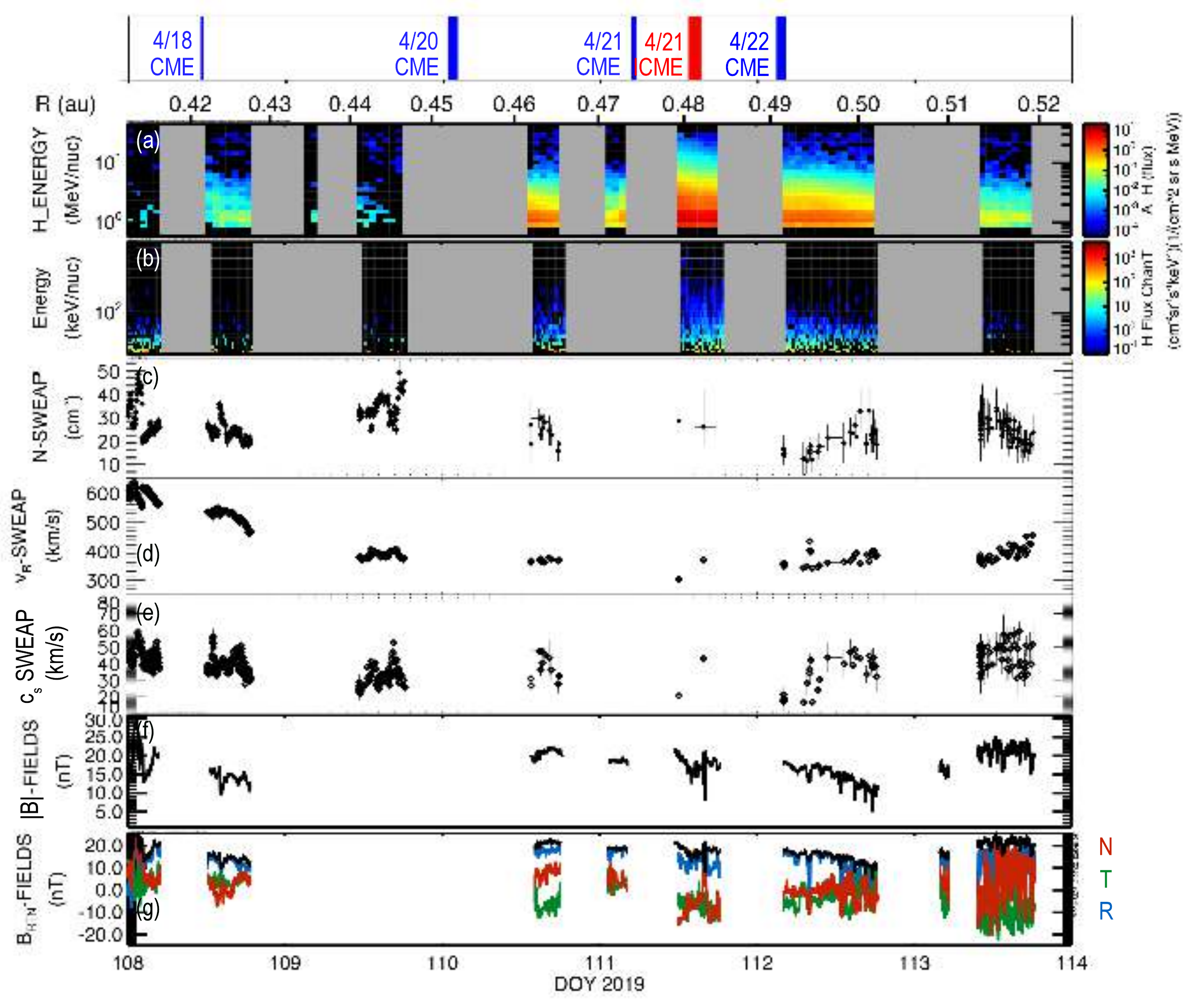} 
\caption{The energetic particle fluxes (outward along the Parker
  spiral) observed by \ISOIS~ are shown together with SWEAP plasma
  and FIELDS observations during the period from April 18 to April 24,
  2019. Panels (a) and (b) show energetic proton differential fluxes
  observed by EPI-Hi and EPI-Lo, respectively, in directions outward
  away from the Sun along the nominal Parker spiral magnetic
  field. Panels (c) -- (g) show the solar wind density, radial wind
  speed, thermal speed, magnetic field strength (black curve in Panels
  f and g), and RTN field components. Here, RTN refers to Radial
  ($\mathbf{r}$, blue curve), Tangential ($\mathbf{t} \propto
  \mathbf{\Omega}_\odot \times\mathbf{r}$, green curve) and Normal
  ($\mathbf{n} = \mathbf{r}\times\mathbf{t}$, red curve) orthonormal
  components with $\Omega_\odot$ defined as the spin-axis of the Sun
  (in J2000).  The two top labels show radial distance and the start
  times of CME events observed by STEREO and LASCO. }
\label{fig:fluxSweap}
\end{figure}

The CME released on April 20 overtook the PSP spacecraft on April 21
near 16:00 UTC when the \ISOIS~ instruments, 
solar wind instruments (SWEAP) and magnetic field instruments
(FIELDS) were powered on. Figure \ref{fig:fluxSweap} shows the
energetic particle data from EPI-Hi and EPI-Lo plotted together with
plasma and field data on PSP.

Using the time difference between the CME release on April 20 and the
observation at PSP on April 21, we infer an average Interplanetary CME
(ICME) speed of $\sim$515 km s$^{-1}$. This average propagation speed
is larger than 387 km s$^{-1}$ speed inferred from observations of
CMEs near the Sun listed in the DONKI database. However, CME
measurement near the Sun remains somewhat subjective
\cite[]{Webb:2012} and coronagraphs identify the propagation of the
the core CME that drives the plasma.  The identification of shocks or compressions in
front of the core CME (often in the form of a flux rope) in white
light coronagraph images has been very difficult
\cite[e.g.,][]{Vourlidas:2003, Ontiveros:2009, Vourlidas:2009}.
Therefore, the larger average propagation speed deduced from the
compression that leads the CME compared to the speed derived from
coronagraph images is expected. The arrival time of 16:00 UTC is
similar to the arrival time at PSP of 19:09 UTC predicted from the
WSA-ENLIL+Cone Model
(https://kauai.ccmc.gsfc.nasa.gov/DONKI/view/WSA-ENLIL/14640/1), which
is detailed in \S\ref{sec:modeling} and Appendix
\ref{sec:enlilApp}.

\begin{figure}
\includegraphics[width=0.95\columnwidth]{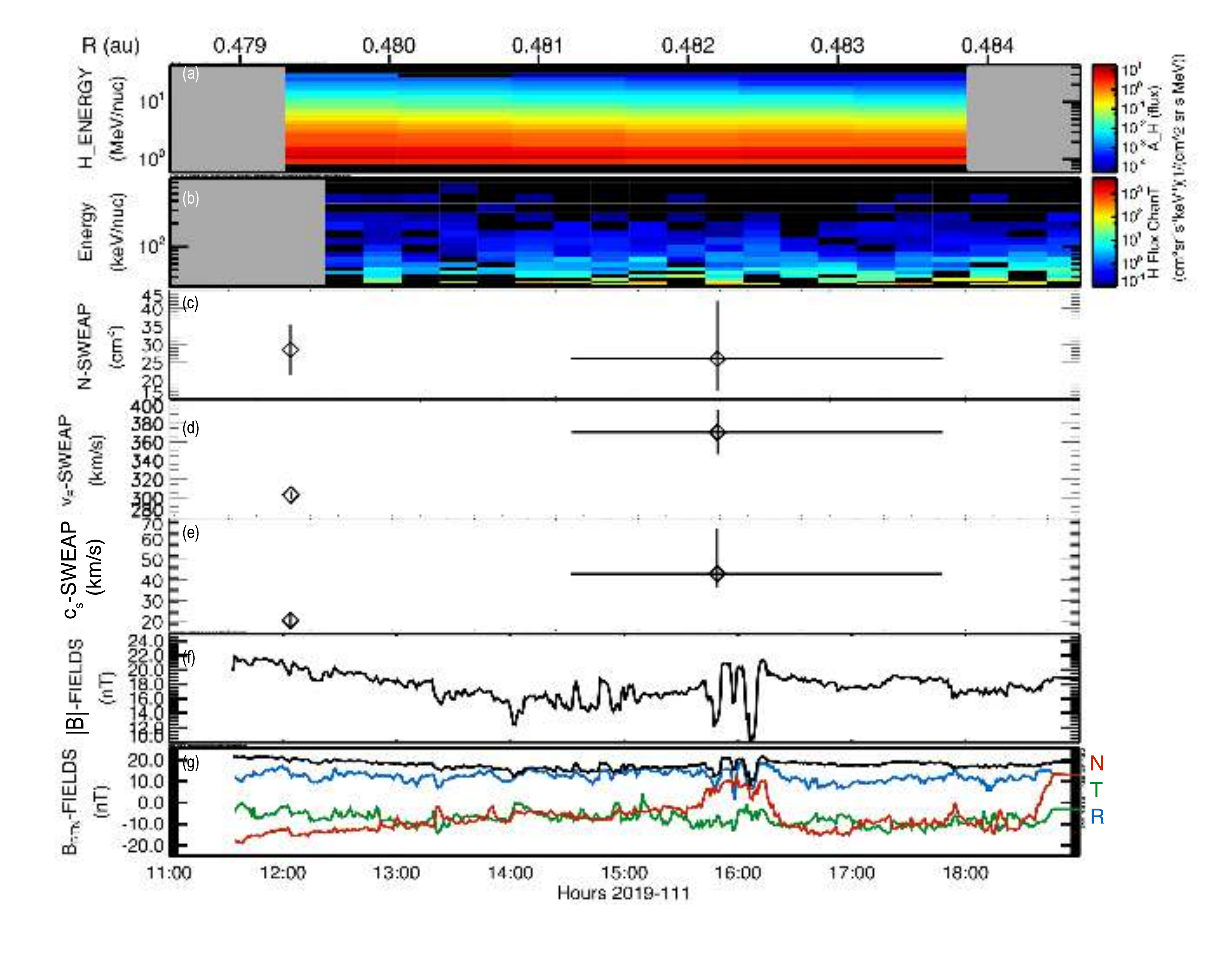} 
\caption{The energetic particle fluxes observed by \ISOIS~ (outward
  along the Parker spiral) together with plasma observations during the
  passage of the ICME on April 21. This ICME was associated with the
  CME released from the Sun on April 20. The average ICME propagation
  speed of 515 km s$^{-1}$ was inferred from the time difference
  between the compression passage at PSP and the CME release. Panels
  (a) and (b) show energetic proton differential fluxes observed by
  EPI-Hi and EPI-Lo in directions outward away from the Sun along the
  nominal Parker spiral magnetic field. Panels (c), (d), (e), (f) and
  (g) show the solar wind density, radial wind speed, thermal speed,
  magnetic field strength (black curve in Panels f and g), and RTN (blue, green and red curves,
  respectively) field components.   The top label shows
  radial distance of the PSP spacecraft. }
\label{fig:fluxExpanded}
\end{figure}

Shown in Figure \ref{fig:fluxExpanded} is an expanded view of the ICME
passage. The compression shows plasma speed rising from $V \sim 300$
km s$^{-1}$ to $V \sim 380$ km s$^{-1}$.  The SWEAP analysis during
this period was a particular challenge since the solar wind signature
in SWEAP was difficult to identify unambiguously. After averaging the
more than two hour time-period used for the second SWEAP data point
centered at about 16:00 UTC on April 21, we were able to identify a
solar wind signature. A gradient in radial solar wind speed is
observed along with a rise in the thermal speed suggesting that a
compressional plasma structure passed by the spacecraft. However, the
lack of time resolution makes it difficult to determine whether the
structure was a shock or a compression. The final speed of the
structure and the associated plasma density are also undetermined. The
average propagation speed of 515 km s$^{-1}$ \revise{was deduced from
  the distance to PSP divided by the propagation time of the CME to
  the PSP spacecraft.  The proagation time was from CME initiation
  observed in coronagraph images to the arrival time at PSP. The
  average 515 km s$^{-1}$ CME speed} and the slow wind speed $\sim$
300 km s$^{-1}$ in front of the ICME suggest that the compression
ratio was of order $V_f/V_s \approx 1.7$, where $V_f$ is the fast wind
speed and $V_s$ is the slow wind speed. This compression ratio must be
taken as a crude estimate given the lack of higher resolution plasma
data. We note from the modeled plasma timeline shown in Appendix
\ref{sec:enlilApp} that the CME may have accelerated as a part of a
larger scale stream interaction region. If this were the case, the CME
compression region forms as a part of an even larger compression
region within the solar wind.

It is possible that the compressed plasma within the ICME is a remnant
of an already merged structure. If this were the case, the ICME plasma
may have been moving more quickly upstream, and subsequently slowed as
the fast flow merged with the slower flow. Therefore, we take a
compression ratio of $\sim$ 1.7 as a lower limit.

As shown in the bottom panel of Figure \ref{fig:fluxExpanded}, the
radial magnetic field (blue curve) and the magnetic field strength
(black curve) are almost equal throughout the April 21 period. The
magnetic field has a large radial component throughout the ICME
passage making it unlikely that a flux rope passed \revise{over} the PSP
spacecraft. The largest $\sim 55^\circ$ deviation of the field from
radial occurs from 16:30 -- 17:30 on April 21.  As discussed in
\S\ref{sec:modeling}, modeling consistently suggests that the PSP
spacecraft was overtaken by the flank of the ICME.

\newcommand{\dg}{$^\circ$}

\begin{deluxetable}{c|cccccc}
\tablecaption{CMEs released from April 18 to 24, 2019\tablenotemark{a} \label{tab:table}}
\tablehead{
\colhead{Date} & \colhead{time} & \colhead{CME \tablenotemark{a}} & \colhead{PSP\tablenotemark{a}}   &  \colhead{Speed }        & \colhead{Width} & \colhead{type III}  \\
\colhead{}     & \colhead{Z}     & \colhead{}    & \colhead{}   & \colhead{(km s$^{-1}$)} & \colhead{($^\circ$)} &   }
\startdata
4/18/2019      & 11:09                & (-149\dg,1\dg)\tablenotemark{b}        & (57\dg,1.5\dg)                  &        428             & 44 & STEREO-A \\
4/20/2019      & 01:25                & (90\dg,2\dg)\tablenotemark{c}           & (60\dg,1.8\dg)                  &        387             & 60 & WIND  \\
4/21/2019      & 05:00                & (117\dg,11\dg)\tablenotemark{d}        &  (61\dg,1.9\dg)                 &        367             & 54 & WIND \\
4/22/2019      & 03:36                & (87\dg,9\dg)\tablenotemark{e}           &  (63\dg, 2\dg)                  &        434             & 52 & WIND\tablenotemark{f} 
\enddata
\tablenotetext{a}{Longitude and Latitude in HEEQ coordinates}
\tablenotetext{b}{The source observed by STEREO A EUVI 195 \AA~ began at 10:05 UTC and was characterized by dimming and opening
field lines along with a post-eruptive arcade. }
\tablenotetext{c}{The eruption from AR 12738 that caused this CME corresponded to a B8.1 flare from the 
active region and a filament eruption visible off the western limb that began at 00:42 UTC. }
\tablenotetext{d}{Initial source is an eruption from AR 12738, just beyond the western limb in AIA 171 and 304, at 03:24 UTC. 
Later eruption visible behind the western limb in AIA 171 and 304 at 05:00 UTC which may have contributed to the later/brighter 
inner edge of ejecta seen in the CME. }
\tablenotetext{e}{The source is an eruption in AR 12738 around 02:50 UTC. 
The CME is very faint and the real time measurements were done while there was a data outage in STA Cor2. 
\tablenotetext{f}{\revise{Very weak type III emissions were observed prior to and during the onset of the CME event on April 22.} }
}
\end{deluxetable}

\section{Contemporaneous Remote and \emph{In Situ} Observations}
\label{sec:electrons}

In assessing the source of seed populations and accelerated particles
observed at PSP, it is important to identify solar flares and Type III
radio bursts observed over the period studied. Energetic electrons are
often nearly scatter-free \cite[e.g.,][]{Lin:1974} and provide an
unambiguous identification of energetic particle seed populations.
Further, the observed Type III emissions over this period provide
remote association with energetic particles release near CME ejection.

\begin{figure}
\includegraphics[width=0.95\columnwidth]{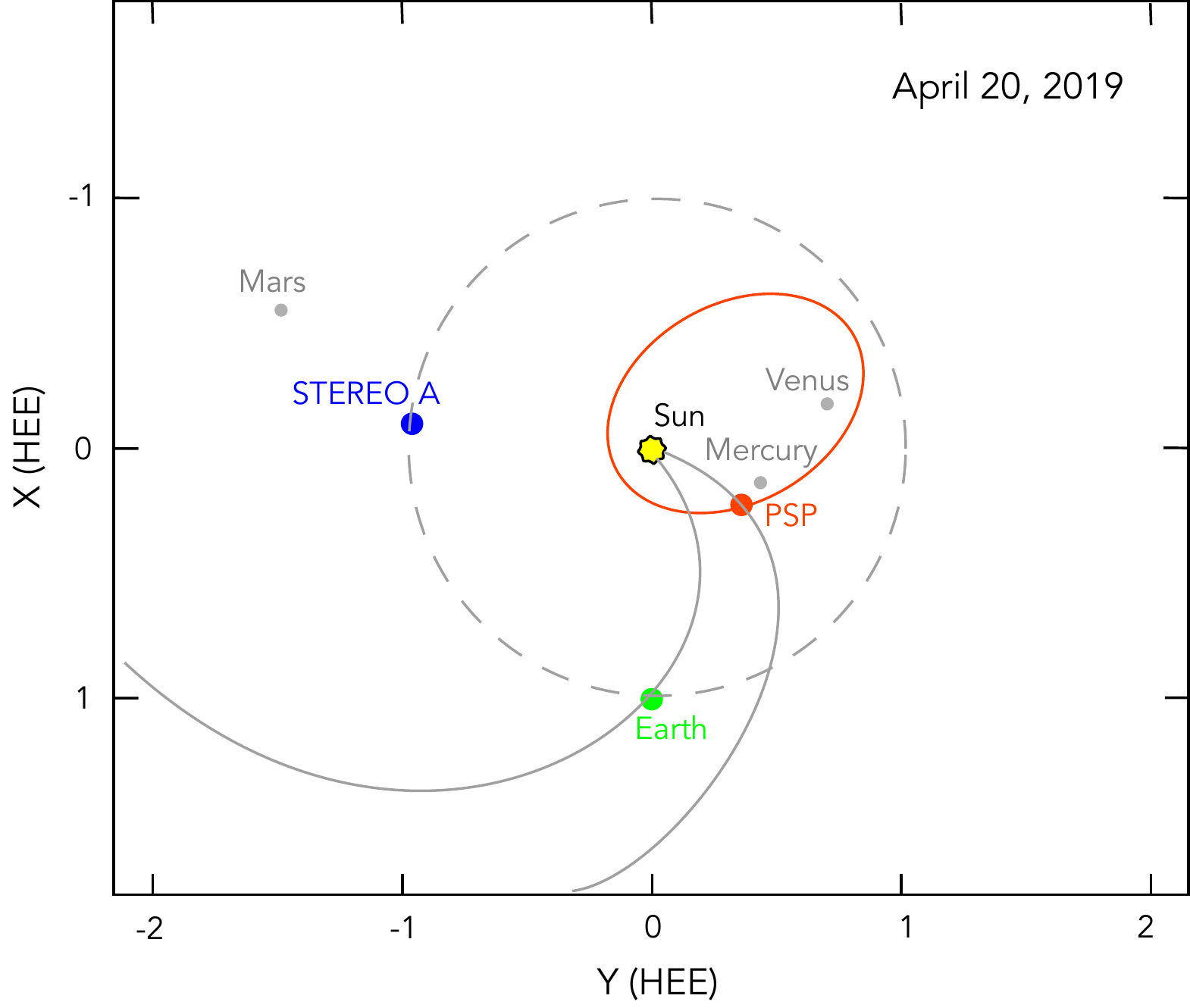} 
\caption{Illustration of the position of PSP in the ecliptic plane on
  April 20, 2019 relative to Earth and other planets within the inner
  heliosphere, and STEREO A.  The red curve shows the second orbit of
  PSP about the Sun, and the solid grey curves show nominal (400 km
  s$^{-1}$) Parker spiral magnetic fields lines connected to PSP and
  to Earth. The dashed grey curve represents the 1 au circle in the
  ecliptic plane. The coordinate system used here is Heliocentric
  Earth Ecliptic (HEE) \cite[]{Russell:1971, Hapgood:1992,
    Franz:2002}. }
\label{fig:orbit}
\end{figure}

The PSP spacecraft was \revise{East
of Earth in the heliocentric frame (i.e., ahead and upstream of Earth in its orbit)}
on Parker spiral field lines less than 25$^\circ$ from
near-Earth spacecraft. Figure \ref{fig:orbit} shows the PSP orbit 
relative to the Earth and other planets within the inner heliosphere,
and relative to the STEREO A spacecraft. This configuration proves to
be important because the  ACE spacecraft located at the
Lagrangian L1 point was close to being magnetically connected to
PSP. As such, we use observations from ACE of energetic electrons as
indicators of seed populations.
 
\begin{figure}
\includegraphics[width=0.95\columnwidth]{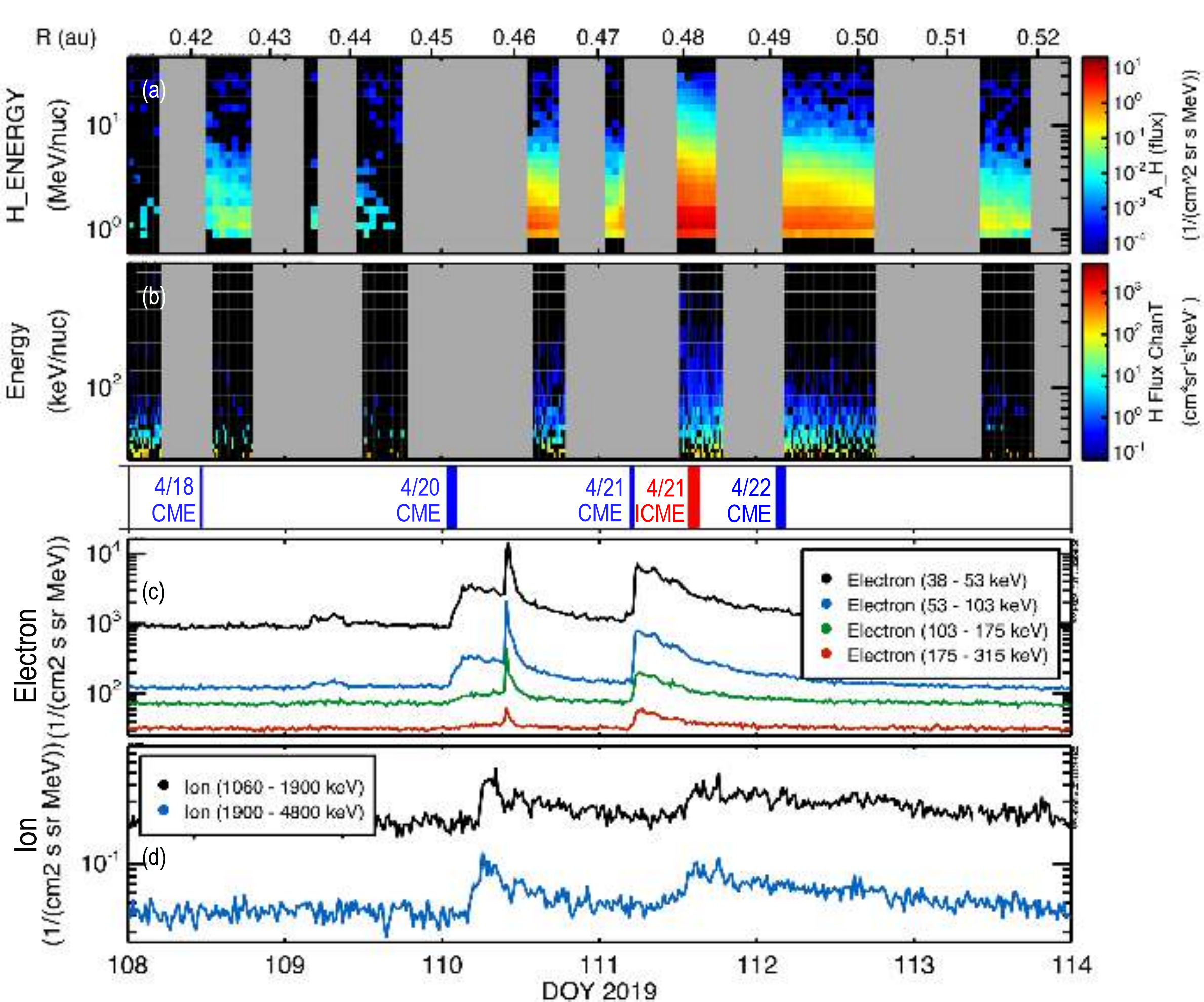} 
\caption{The energetic particle fluxes (panels a and b, outward along
  the Parker spiral) together with spin and 20-minute averaged
  intensities of 38-315 keV electrons versus time, observed with the
  ACE/EPAM experiment in the interval April 18-24, 2019 (panel
  c). 1.1-4.8 MeV spin and 20-min averaged ion intensities observed
  with the LEMS120 detector of the ACE/EPAM experiment (panel d).  The
  middle label CMEs identified in Table 1 and the top label shows
  radial distance of the PSP spacecraft. }
\label{fig:fluxElectrons}
\end{figure}

Figure \ref{fig:fluxElectrons} Panel (c) shows an overview of the
20-min and spin-averaged differential intensities of four electron
channels (DE1-DE4: 38-315) as measured by the B detector head of the
CA60 telescope of the EPAM (Electron, Proton and Alpha Monitor)
experiment on ACE. Details regarding the instrument and observations
over the period studied are provided by Appendix A. Panel (d) shows
20-min and spin-averaged differential intensities of 1.1-4.9 MeV
energetic ions measured by the LEMS120 telescope of the EPAM
experiment for the same interval.

On April 20 and 21, two near-relativistic prompt electron events are
observed (E1 and E2). The electron event on April 21 (E2) is more
intense and extends to a higher energy range, up to 315 keV. An
electron event is observed superposed during the decay phase of the
first event of this period. The electron events exhibit typical
rise-time to maximum of a few tens of minutes, a long smooth decay
\cite[]{Lin:1970, Lin:1974}, and beam-like pitch-angle distributions,
as detailed in Appendix A. In these prompt electron events, particles
accelerated in a magnetically well-connected solar source region
arrive abruptly at the spacecraft \cite[e.g.][]{Reames:1999,
  Malandraki:2002}. Two, relatively weak, proton intensity
enhancements (Figure \ref{fig:fluxElectrons}, Panel d) are observed in
association with these electron events.

Appendix A discusses observed electron pitch-angle distributions used
to estimate the event onset times at the Sun: E1 at 00:44 UTC on April
20, roughly 40 minutes prior to the corresponding CME release time
listed in Table \ref{tab:table}; and E2 at 04:47 UTC on April 21,
2019, which is 13 minutes prior to the CME release time listed in
Table \ref{tab:table}.  
 
Both events on April 20 and April 21 showed strong enhancements in
$^3$He \cite[]{Wiedenbeck:2019} observed by \ISOIS. These strong
enhancements definitively support the concept that energetic particle
seed populations contained flare-accelerated material
\revise{\cite[][]{Mason86,Reames:1999,Mason:2002,Desai:2003,Bucik:2015,Bucik:2016}}.

The CMEs initiated on April 18, April 20 and April 21 (Table
\ref{tab:table}) were associated with Type III bursts observed by
STEREO A Radio and Plasma Wave Investigation (WAVES) on April 18, and
by the Wind Radio and Plasma Wave Investigation (WAVES) on April 20
and April 21\footnote{Browse data for these experiments are available
  at the Goddard Space flight Center STEREO Science Center,
  https://stereo.gsfc.nasa.gov/browse/.}. There were \revise{only
  weak} Type III emissions observed by Wind/WAVES on April 22
\revise{prior to and during the CME initiation}. Type III bursts start
at around 10 MHz and then progress to lower frequencies with
time. These bursts are delayed with respect to the associated flare
and last on average $\sim$ 20 minutes.  Type III bursts are associated
with CMEs and typically solar energetic protons \cite[]{Cane:2002,
  MacDowall:2003}.  However, \cite{Gopalswamy:2010} found that a Type
III burst does not always signify the presence of solar energetic
protons.

\section{Modeling of the April 20-21 SEP event at PSP}
\label{sec:modeling}

The energetic particle events studied were
associated with CMEs, suggesting a
relationship with a solar active region. The PSP spacecraft was at 
heliocentric distances ranging from 0.46 to 0.49 au during the April~20--22 time
period. The SEP events detected by PSP were mapped along the Parker
spiral back to the Sun using the Current Sheet Source Surface (CSSS)
Model, detailed in Appendix \ref{sec:CSSS}.


In Figure \ref{fig:backmapping}, the black dotted line
represents the solar equator and the symbols (triangles) 
indicate the footpoint locations on the
source surface at 15 R$_s$ in the corona mapped back on April 20, 21 and
22. The filled circles show
the respective photospheric footpoints mapped back along the open
magnetic field lines. Note that the photospheric footpoints lie close
to the active region (AR12738) boundaries.

\begin{figure}
\includegraphics[width=0.95\columnwidth]{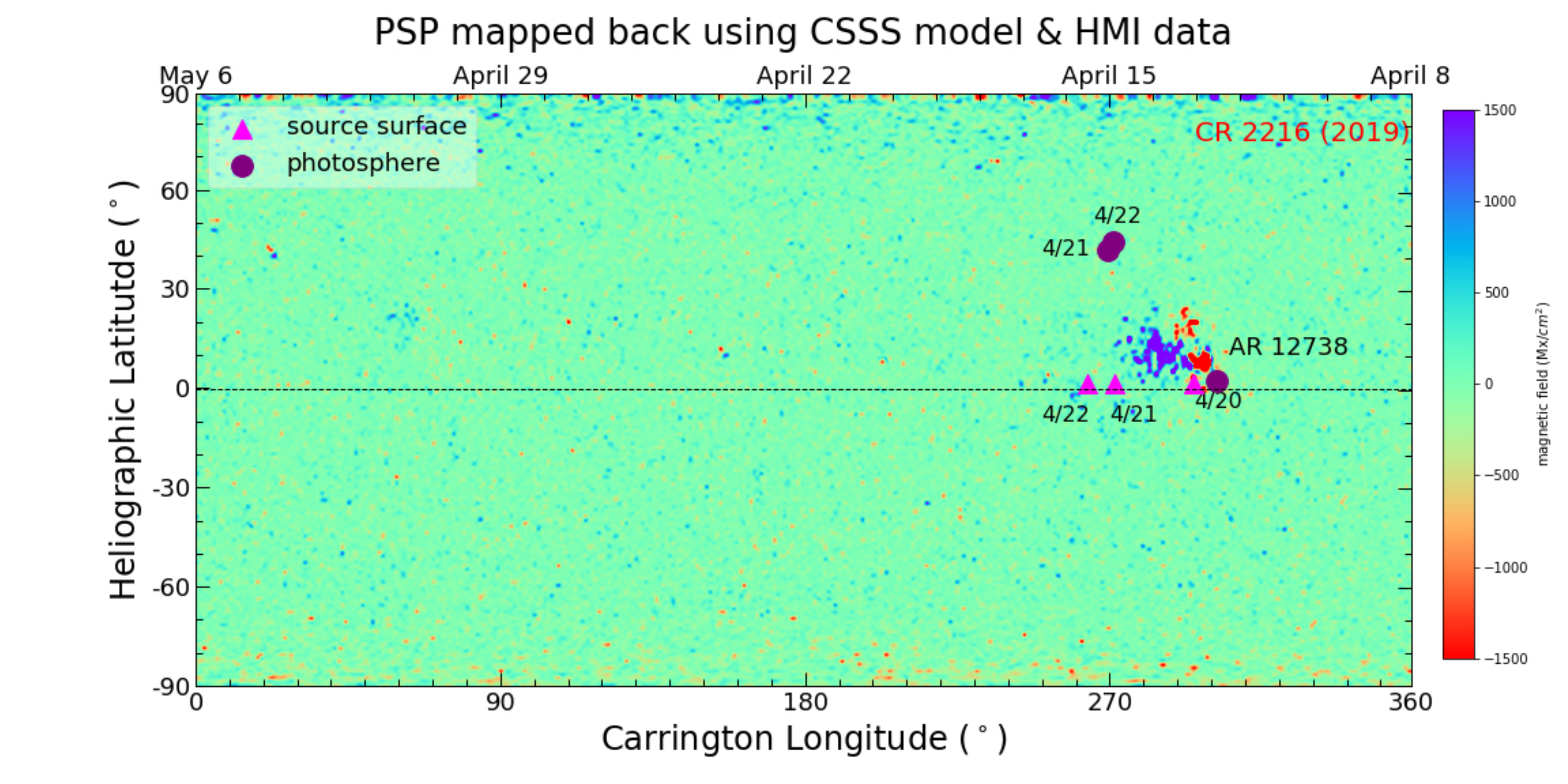} 
\caption{Source surface model to map magnetic fields from the PSP
  location to source regions at the Sun. The triangles are the angular
  positions of PSP mapped back to the source surface at 15 R$_s$ and
  the circles show the corresponding footpoints at the photosphere on
  April 20, 21, and 22. These footpoints were back-mapped using solar
  wind speeds of 300, 350, 400 km s$^{-1}$, respectively.  Footpoints
  are over plotted on the HMI synoptic map for Carrington rotation CR
  2216 which includes the SEP event observed by PSP during April 20 --
  22.
 }
\label{fig:backmapping}
\end{figure}

\begin{figure}
\includegraphics[width=0.95\columnwidth]{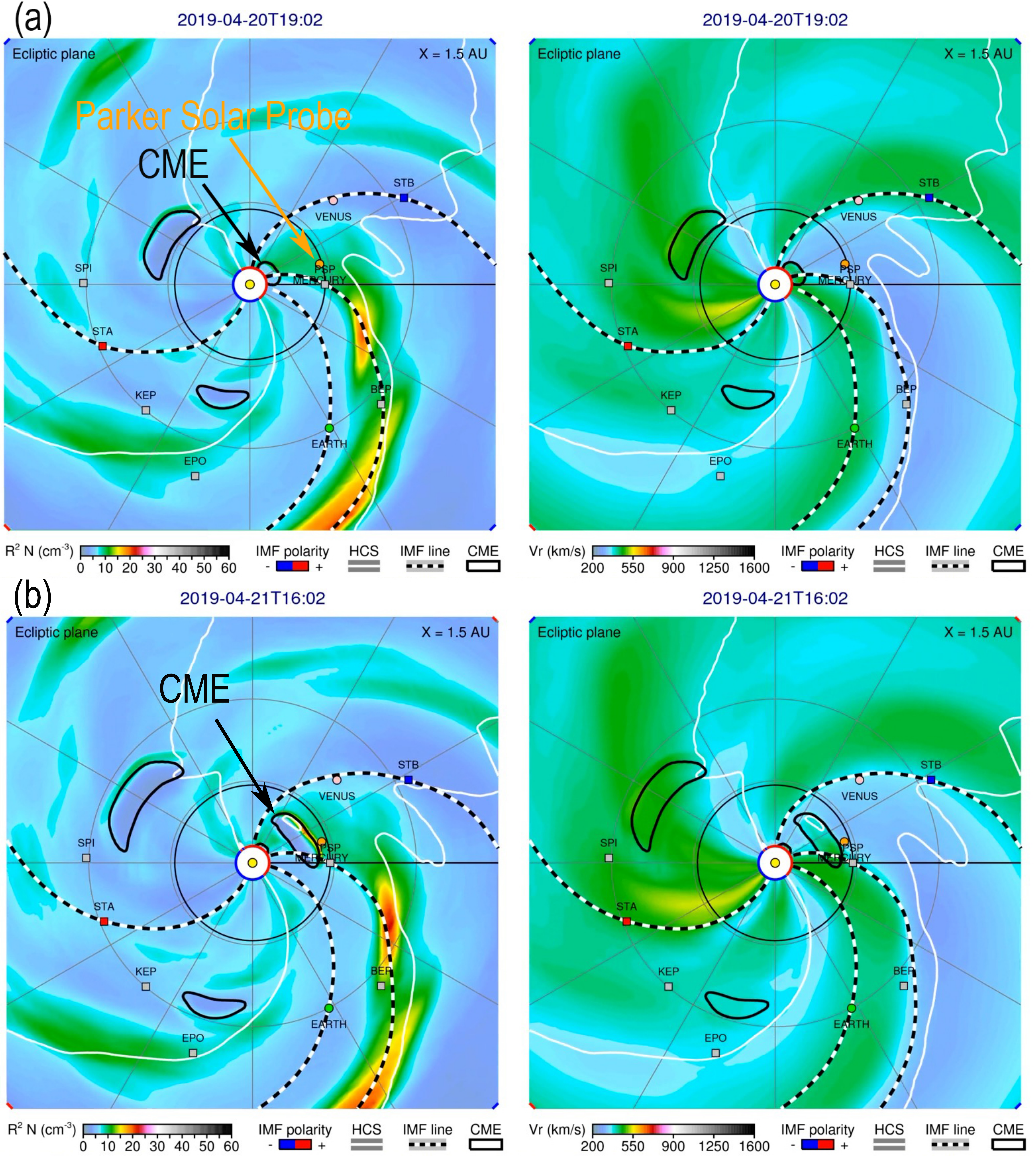} 
\caption{Snapshots showing the Enlil model of the April 20, 2019 CME
  released at 01:25 UTC and propagating out to PSP. Panels (a) - (b)
  show the evolution of the CME through the inner heliosphere at
  different stages of the CME's propagation to PSP. Left panels show
  modeled densities and right panels show the speed structure from the 3-D model. 
}
\label{fig:enlil1}
\end{figure}

\begin{figure}
\includegraphics[width=0.95\columnwidth]{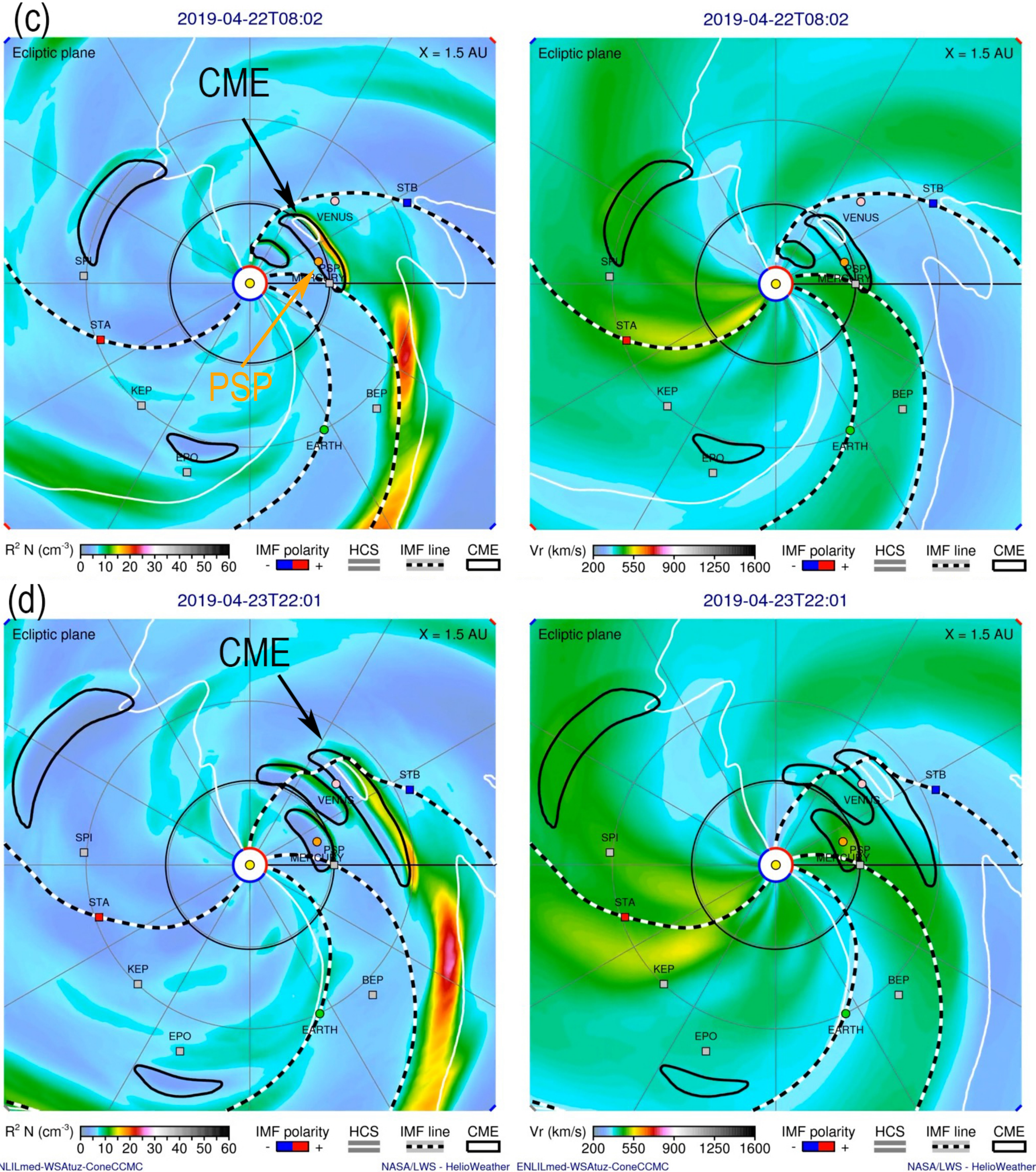} 
\caption{Enlil model snapshots similar to Figure \ref{fig:enlil1}
  showing the stages of the CME's propagation after April 21 out to 1
  au.}
\label{fig:enlil2}
\end{figure}

Figures \ref{fig:enlil1} and \ref{fig:enlil2} show the propagation of
the CME released on April 20 from the Sun using the Enlil model
\cite[]{Odstrcil:2003} initialized using CME parameters from the
CCMC's DONKI database. The location of PSP is on the flank of the CME,
which is consistent with plasma compression during the passage of the
ICME without the accompanying signature of a flux rope. Further
results from Enlil modeling are detailed in Appendix
\ref{sec:enlilApp} and supplementary online materials include a movie
of the CME.

The compression ratio, estimated to be $r_c = 1.7$, is difficult to
determine because SWEAP observations were incomplete during the ICME
passage and the plasma speed and density were likely time variable, as
suggested by the results of the Enlil model (Figures \ref{fig:enlil1}
and \ref{fig:enlil2}).  This compression ratio is similar to that
found from the Enlil simulation.

\begin{figure}
\includegraphics[width=0.95\columnwidth]{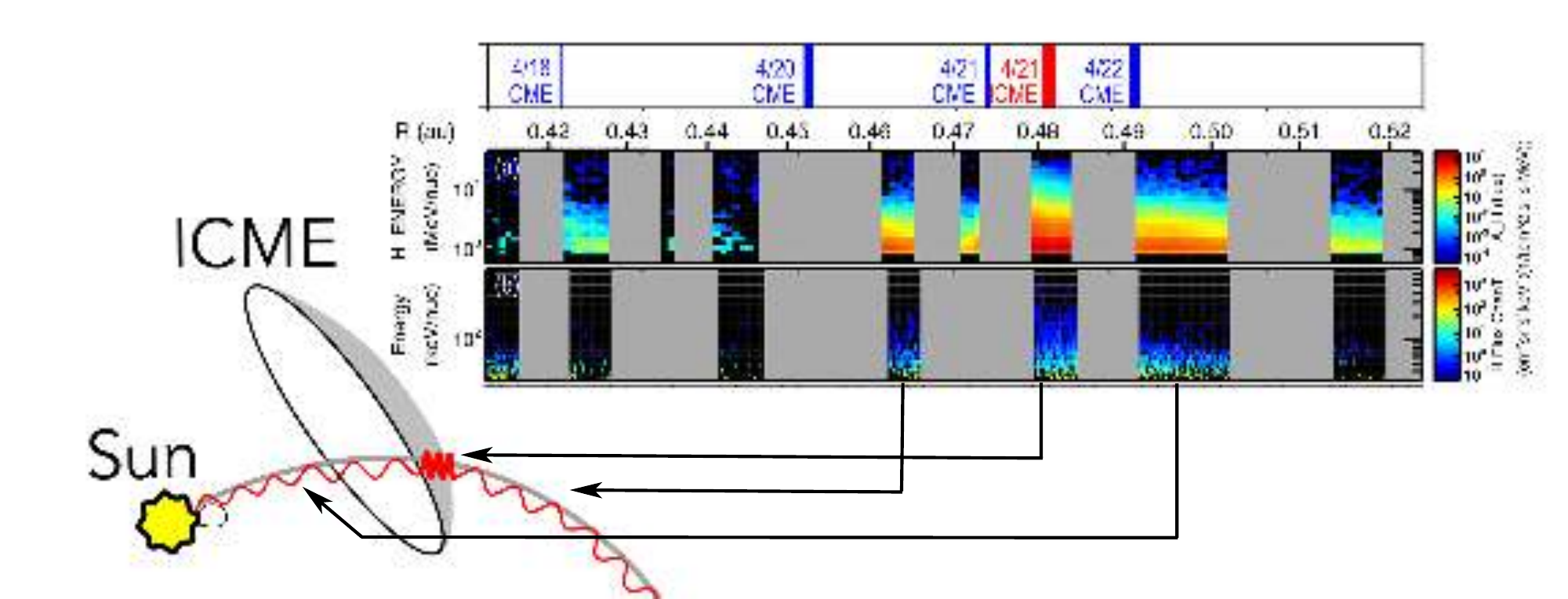}
\caption{Schematic of the the Interplanetary Coronal Mass Ejection
  (ICME), which drives a compression region (shown by grey region)
  ahead of the ICME. We show the fluxes observed by \ISOIS~ before,
  during and after the ICME passage, and indicate where relative to
  the ICME these populations are observed.  }
\label{fig:ICMEcartoon}
\end{figure}

The $^3$He enhancements observed during this period
\cite[]{Wiedenbeck:2019} confirm that solar flares are responsible
  for producing at least part of the energetic particle seed population
  observed. Appendix \ref{sec:compress} considers the situation where
  an energetic particle population is swept up, compressed and
  accelerated with the solar wind plasma in front of the ICME, as
  depicted in Figure \ref{fig:ICMEcartoon}.

Appendix \ref{sec:compress} shows that if the width of the compression
is wider than the diffusion region, particle distributions 
are convected through the compression and accelerated by the solar
wind speed gradient. As a result, the distribution function in the
compression is given by $f_c = r_c^{\gamma/3} \tilde{f}_0$ where $r_c$
is the compression ratio ($r_c = 1.7$), and $\gamma$ is the power-law index of
the energetic particle distribution $\tilde{f}_0$ in the faster wind behind the
compression such that $\tilde{f}_0 \propto p^{-\gamma}$. The criterion required
for this form of acceleration is that the scattering mean free path
$\lambda$ must be sufficiently small compared to the width, $\delta x$,
of the speed gradient to restrict diffusion upstream. Therefore,
$\lambda < 3 u \delta x / v$ where $u$ and $v$ are the solar wind and
particle speeds, respectively.  We estimate the compression region
width of $\sim 0.07$ au based on an average convection speed of 400 km
s$^{-1}$ and the time period of $\sim 7$ hrs over which the enhanced
energetic particle fluxes are observed (see Figure
\ref{fig:fluxExpanded}). For these parameters, the mean free path of a
1 MeV proton would have to be less than 0.006 au or $\sim 1.3$ R$_s$
to restrict upstream diffusion. 

For reference, we show the modeled time profile of the event at PSP in
Appendix \ref{sec:enlilApp}.  The modeled compression region is more
than 0.25 au in scale, which is larger than the the inference using
\ISOIS~ data. A significant portion of the compression region
appears to be missed when the instruments were powered off over
satellite contacts and high-speed data transfers. This potentially
larger size of the compression region would slightly relax the limit
on the scattering mean free path, $\lambda < 0.02$ au. More
importantly, the model results indicate that the ICME may be in the
process of being enveloped in a stream interaction region. 

Figure \ref{fig:flux} shows the energetic proton differential fluxes
observed by EPI-Hi and EPI-Lo directed from and to the Sun on the
nominal direction of Parker field lines. These observations show that
the distributions are closest to being isotropic at the highest
energies observed and we observe an outward anisotropy in the
spacecraft reference frame, which is most pronounced at the lowest
energies observed. The observations are consistent with distributions
that are close to being isotropic in the solar wind reference
frame. Standard diffusion theory
\cite[e.g.,][]{Forman:1975} specifies the energetic particle
anisotropy (the Compton-Getting term) in the spacecraft reference frame,
$\mathbf{\xi} \approx \gamma \mathbf{u} /v$. Below 1 MeV, observations
within the compression indicate $\gamma \approx 5.4$. At 100 keV, the
inferred anisotropy magnitude is $\sim 50$ \% and at 1 MeV, the
anisotropy magnitude drops to 16\%. Both the outward direction of the
observed anisotropy, and its magnitude appear roughly consistent with
the Compton-Getting term. 

In Figure \ref{fig:spectra2}, we show the energetic particle
distributions (red curves) that result from the compressive acceleration
of seed populations. For energies below 1 MeV, we use the energetic
particle fluxes ahead of the CME (black data points) as a proxy for
the uncompressed seed population. Above 1 MeV, the solid red curve
shows the prediction using the energetic particle fluxes ahead of the
ICME (black points) as the proxy for the seed population, and with
power-law indices ($\gamma$) based on the two fits indicated in the
Figure ($J \propto E^{-2.8}$ for $1$ MeV $< E \leq 3$ MeV and $J
\propto E^{-4.2}$ for $E > 3$ MeV). The dashed red curve shows the
prediction using the energetic particle fluxes behind the ICME (blue
points) as the proxy for the seed population.

There is a fundamental change in energetic particle fluxes between
energies below and above 1 MeV. The fluxes ahead of the ICME provide
better proxies for the uncompressed seed population below 1 MeV, but
the fluxes of seed populations were likely time variable over the
period observed. For example, it is likely that additional seed
populations were associated with the April 21 04:47 UTC CME. These
additional fluxes may account for the change in the seed population
above 1 MeV. This scenario is consistent provided that slower protons
from the April 21 flare with energies $<$ 1 MeV did not propagate to
PSP during the ICME passage, whereas faster $>1$ MeV protons were
capable of propagating to the spacecraft. 

\revise{ The propagation of seed populations following a delta
  function injection at the Sun creates a spatially and temporally
  dependent variation of the seed particle distribution function
  \cite[e.g.,][]{Schwadron:1994}. Early in an event, particles move
  out from a flare at the causal limit dictated largely by the
  particle speed, and later in events distributions relax into
  diffusive propagation. A 1 MeV proton has a speed of $v =
  $13.8$\times 10^3$ km s$^{-1}$. We take a propagation distance of
  $\delta x = 0.48$ au to PSP and a propagation time of $\delta t = $
  11.2 hr from the point of energetic particle injection to the
  observation time at PSP. The causal limit for propagation is at a
  distance of 3.7 au. Because PSP is so much closer in at 0.48 au, we
  can safely consider distributions that evolve diffusively, with a
  spatial profile $f_D \propto \exp\left(-x^2/[4 \kappa_\parallel t]
  \right)$. We take the parallel diffusion coefficient given by
  $\kappa_\parallel = \lambda v/3$ where $\lambda$ is the scattering
  mean free path. We use the proxy that the characteristic diffusive
  propagation distance is where the spatial profile falls to a factor
  of 2 lower than near the source, allowing us to estimate the one
  unknown, the scattering mean free path, $\lambda = 3 \delta x^2/ (4
  \ln(2) v \delta t) \approx 0.1$ au.  } This estimate appears roughly
consistent with scattering mean free paths observed previously over
similar ranges of rigidity \cite[]{Bieber:1994}; however the
assumption that the mean free path is independent of distance can be
questioned. For example, if we assume that the scattering mean free
path scales with the radial distance, then we would infer a slightly
\revise{larger average mean free path, $\sim 0.11$ au, than that inferred
  with} no radial dependence .

Our results show that compressive acceleration of seed populations
requires a scattering mean free path smaller than $\sim 0.006$ au,
\revise{which is about 17 times smaller than the $\sim 0.1$ au mean free path
estimated for energetic particle propagation from the April 21 CME
ejection.} In other words, compressive acceleration of seed populations
requires a reduced scattering mean free path within the compression
region. It is possible that this reduced mean free path is the natural
outcome of compression, or that the CME itself plays a role in
reducing the scattering mean free path.  

The reduced mean free path \revise{in the ICME compression} is
also consistent with an instability driven by the streaming and
subsequent scattering of high fluxes of energetic protons
\cite[]{Stix:1992, Melrose:1980}. The instability was invoked in
models of wave growth and diffusive shock acceleration
\cite[]{Lee:1983}.  Observations show that 3 -- 6 MeV proton
intensities early in large gradual events did not exceed a plateau
value of $\sim$ 100 -- 200 (cm$^2$ sr s MeV)$^{-1}$
\cite[]{Reames:1990}, which was subsequently dubbed the ``streaming
limit,'' although intensities could rise much higher during passage of
the shock. Observations \cite[]{Ng:1994} show that wave growth greatly
limits the flow and streaming of protons. Although the differential
fluxes observed in this paper are significantly lower than the
streaming limit, the reduced scattering mean free path within the
compression appears generally consistent with streaming limited
scenarios.  Observations were also used to extend the streaming
limit to higher energies \cite[]{Reames:1998} and showed how the
low-energy spectra can be flattened, but only when sufficient
intensities of high energy protons precede them
\cite[]{Reames:2010}. Therefore, the presence of high fluxes of seed
populations preceding the events observed may be critical to limiting
the scattering mean free path throughout the compression region.

\revise{Given the reduced scattering mean free path of 0.006 au in the
  compression region, we estimate in Appendix \ref{sec:compress} that
  the time required for diffusive acceleration to 1 MeV requires more
  than 2.5 days. This result demonstrates that even with a reduced
  mean free path within the compression region, the local particle
  acceleration rate to 1 MeV is still too low to account for the
  changes in fluxes observed throughout the event. This low
  acceleration rate therefore reinforces the need for pre-existing
  seed populations that are fed the compression region, and
  demonstrates why the energetic particle spectrum within the
  compression region remains so similar compared to the differential
  energy fluxes up- and downstream from the compression.  }

\revise{Given the importance of pre-existing seed populations needed
  to explain the observations within the ICME compression, it is
  important to ask further how these seed populations were
  generated.} The results obtained from the changes to differential
energy spectra shown in Figure \ref{fig:spectra2} indicate that
compression of seed populations within the solar wind plasma accounts
for the increase in differential energy fluxes during the passage of
the ICME. However, the question remains as to how the seed populations
are produced. As discussed previously, the presence of enhanced $^3$He
throughout the events provides definitive evidence that flares
contribute to the seed populations observed
\cite[][]{Mason86,Reames:1999,Mason:2002,Desai:2003}. Curiously
though, the compression ratio of $r_c \approx 1.7$ inferred for ICME
compression would yield a differential energy spectrum with a
power-law of $E^{-2.6}$ based on DSA. The power-law above 1 MeV is
$\sim E^{-2.8}$, similar to the DSA prediction, and we observe a
steeper power-law of $\sim E^{-4.2}$ above $\sim 3$ MeV. This
characteristic broken power-law distribution was described by
\cite{Schwadron:2015SEP} and \cite{Schwadron:2015JPCS} as a product of
particle acceleration driven by CMEs in the low corona. Alternatively,
\cite{Li:2009} demonstrated that such broken power-laws may also
naturally result from quasi-perpendicular shocks.


Another feature observed in the spectrum below 1 MeV is a differential
flux power-law of $E^{-1.7}$. Based on the diffusion calculation
presented previously, it appears that the $< 1$ MeV protons are
sufficiently \emph{immobile} so that they cannot propagate directly
from the April 21 CME and flare source to PSP during the ICME passage. In
other words, the $<1$ MeV particles are likely to interact over longer
propagation periods within the solar wind plasma prior to being swept
up by the ICME compression. It may not be surprising then that these
particles exhibit a harder spectrum. In fact, the spectrum is so hard
that it is close to the $E^{-1.5}$ limit of possible stationary state
plasma distributions out of equilibrium
\cite[]{Livadiotis:2009, Livadiotis:2010}. The seed population below 1
MeV is likely a superposition of particles from multiple flares and
compressions in the solar wind.  \cite{Schwadron:2010h}
argued that superposed distributions are a natural source of
kappa-distributions with hard suprathermal power-laws in the typical
range of $E^{-2.5}$ - $E^{-1.5}$. The $E^{-1.5}$ spectrum was also
considered  a ``ubiquitous'' characteristic of
the low-energy seed population \cite[]{Fisk:2006} within the solar
wind.  A pump mechanism detailed by \cite{Fisk:2008} and
\cite{Fisk:2010} to account for this $E^{-1.5}$ spectrum remains
controversial \cite[]{Jokipii:2010}.

We conclude this section by noting the significant differences between
the \ISOIS~ observations from April 18--24, 2019 compared to the
observations on Nov 11, 2018 \cite[]{Giacalone:2019} of solar
energetic particles produced by a slow coronal mass ejection when PSP
was at $\sim$0.25 au. The particle event showed velocity dispersion
with higher energy protons arriving well before the lower energy
ones. After onset, the particle intensities increased gradually over a
period of a few hours, reaching a peak, and then decayed gradually
before the arrival of the CME at PSP. The SEP intensity decreased
significantly when the CME crossed PSP. The differential energy
spectrum was nearly a power-law as a function of energy with a soft
$E^{-4.73}$ spectrum (40-200 keV). By comparison, the spectral slope
in the April 21, 2019 event below 1 MeV was much harder, $E^{-1.7}$, but
the higher energy slope above 3 MeV was also quite soft, $E^{-4.2}$.

During the Nov 11, 2018 event, anisotropies show that the earliest
arriving particles moved radially outward from the Sun along the
interplanetary magnetic field. However, later in the event the
observed anisotropies are consistent with the advection of an
isotropic distribution. This behavior is consistent with the
observations throughout the April 21, 2019 event, indicating
significant interplanetary scattering of the energetic particles.

\section{Summary and Conclusions}
\label{sec:conclusion}

We have investigated energetic particles observed by \ISOIS~ from April
18 through April 20 of 2019. During this period the Sun released
multiple CMEs, three of which propagated out relatively near
PSP. This period was unique because the PSP spacecraft was close
to being magnetically connected to spacecraft near Earth. The
vantage point of PSP in the inner heliosphere at $\sim 0.5$ au during
a period of generally low activity provided us with the ability to
observe relatively isolated CME events and their interaction with
energetic particle seed populations.

We observed a time period on April 21 in which the flank of an ICME
passed over the PSP spacecraft. The solar wind plasma ahead of the
ICME contained strong enhancements in energetic particle fluxes that
appear to have been compressed and accelerated within the sheath ahead
of the CME. The contemporaneous observation of strong
$^3$He enhancements \cite[]{Wiedenbeck:2019} confirm that seed
populations are rich in material released by solar
flares. Back-mapping of the solar wind's magnetic field on April 21
place the fieldline footpoints close to the boundaries of Active
Region 12738.  Each of the events observed on April 18, 20, and 21
were also associated with the release of CMEs from the Sun. And, on
April 21, we observed a broken-power law above 1 MeV in the compressed
ICME sheath consistent with predictions of diffusive shock
acceleration from shocks or compressions from low in the corona.

\revise{ Observations show that the passage of the ICME on April 21
  was associated with an abrupt increase in energetic particle fluxes
  (see Figures \ref{fig:flux} and \ref{fig:spectra2}).  This abrupt
  increase is inconsistent with diffusive shock acceleration, which
  invokes a diffusive ramp that increases exponentially to the shock
  interface (Equation \ref{eq:diff1}). The abrupt increase in
  differential fluxes is observed together with the lack of a clearly
  defined shock, the general presence of seed populations upstream of
  the ICME, and a differential energy spectrum that approximately
  maintains its form before, during and after ICME passage. Together
  these observations suggest that compression in front of the ICME
  also enhances the fluxes of energetic particle seed populations. }

\revise{In Appendix \ref{sec:compress} we discuss a compression
  mechanism to explain the enhanced energetic particle fluxes ahead of
  the ICME.  The mechanism has no free parameters: the compression
  ratio $r_c$, and seed population spectral index $\gamma$ determine
  the enhancement in the energetic particle flux within the
  compression region (Equation \ref{eq:sol}).  The fluxes predicted by
  the mechanism are generally similar to observations
  within the ICME-driven compression (Figure \ref{fig:spectra2}).  }

\revise{ The local enhancement of energetic particle seed populations
  requires restricted propagation within the compression. If this
  restricted propagation is caused by increased scattering, the mean
  free path of a 1 MeV proton would have to be less than 0.006 au or
  $\sim 1.3$ R$_s$ within the compression (see \S\ref{sec:modeling}).}

\revise{In comparing the seed population before and after the ICME
  passage in Figure \ref{fig:spectra2}, we observe some increase in
  energetic particle fluxes above 1 MeV after the ICME passage.  These
  increased fluxes were likely associated with energetic particles and
  flare particles associated with the April 21 04:47 UTC CME, as
  detailed in \S\ref{sec:modeling}. The timing of the changes in the
  $>1$ MeV seed population fluxes indicate a scattering mean free path
  for the energetic particle seed populations of $\sim$ 0.1 au, more
  than a decade larger than the $\sim$ 0.006 au mean free path needed
  to restrict seed populations within the ICME compression.  It is
  possible that the restricted propagation within the compression
  region is the natural outcome of compressed plasma, or that the
  ICME itself plays a role in reducing the scattering mean free path.
  Wave growth may greatly limit the flow and streaming of protons
  \cite[]{Ng:1994} and the observed restricted propagation has
  physical similarities to observations of streaming limited energetic
  particle fluxes \cite[]{Reames:1990}. Future theoretical work is
  needed to develop a deeper understanding of the restricted
  propagation within the ICME compression. }

\revise{With a reduced scattering mean free path of 0.006 au in the
  ICME compression region, the diffusive shock acceleration to 1 MeV
  would require more than 2.5 days. It is therefore unlikely that the
  local enhancements observed on April 21 at PSP can be accounted for
  by local diffusive shock acceleration.  Consistently, the observed energy spectrum
  within the ICME compression in Figure \ref{fig:spectra2} does not
  show significant changes in the differential energy spectrum beyond 
  the increased fluxes at all energies observed.   }

\revise{A break in the differential energy spectrum at $\sim$ 3 MeV is
  observed throughout the observed events (see Figure
  \ref{fig:spectra2}).  The compression ratio of $r_c \approx 1.7$
  inferred for the ICME compression would yield a differential energy
  spectrum with a power-law of $E^{-2.6}$ based on diffusive shock
  acceleration. The power-law above 1 MeV is $\sim E^{-2.8}$, similar
  to this prediction, and we observe a steeper power-law of $\sim
  E^{-4.2}$ above $\sim 3$ MeV. This characteristic broken power-law
  distribution was described by \cite{Schwadron:2015SEP} and
  \cite{Schwadron:2015JPCS} as a product of particle acceleration
  driven by CMEs \emph{in the low corona}. \cite{Li:2009} found that
  these broken power-laws result from quasi-perpendicular
  shocks. These particle acceleration scenarios for the seed
  population would require the formation of stronger shocks or
  compressions, and much smaller scattering mean free paths in the
  strong magnetic fields low in the corona near these shocks and
  compressions for rapid particle acceleration.  }

\revise{We return to the open question} as to how the energetic
particle seed populations are fed into particle acceleration at
interplanetary shocks. It has been widely known that energetic
particle seed populations are often rich with nearly scatter-free
electrons and species such as $^3$He known to be flare associated
\cite[]{Mason86,Reames:1999,Mason:2002,Desai:2003}.  The 
enhancements in energetic particle seed populations observed in this
study demonstrate how the early evolution of ICMEs \revise{could}
enhance the fluxes of energetic particle seed populations, which
precondition the particle acceleration process at distances further
from the Sun where compressions can steepen into shocks.

The \ISOIS~ observations below 1 MeV show a very hard $E^{-1.7}$ energy
spectrum that is likely a superposition of particles from multiple
flares and compressions in the solar wind. The spectrum is so hard
that it is close to the $E^{-1.5}$ limit of possible stationary state
plasma distributions out of equilibrium \cite[]{Livadiotis:2009,
  Livadiotis:2010}, which suggests that suprathermal particles may play a
more fundamental role for the pressure and heating of the solar wind.


The SEP acceleration process relies on solar flares to produce
energetic particle seed populations, and the acceleration of seed
populations by compressions and shocks driven by CMEs as these
structures propagate through the interplanetary medium. Parker Solar
Probe was at the right place and at the right time to observe the
compression of energetic particle seed populations. Thus, we have
observed a key part of the pre-acceleration process that occurs close
to the Sun in the development of energetic particle events.  The
\revise{enhancement} of energetic particle seed populations observed here
\revise{within the CME-driven compression}
could pre-condition the production of larger fluxes of higher energy
accelerated particles as the compression region grows and steepens
further out in the heliosphere.

\acknowledgements
We are deeply indebted to everyone that helped make
the Parker Solar Probe (PSP) mission possible. We thank all of the
outstanding scientists, engineers, technicians, and administrative
support people across all of the \ISOIS, FIELDS and SWEAP institutions
that produced and supported the \ISOIS, FIELDS and SWEAP instrument
suites, support its operations and the scientific analysis of its
data. This work was supported as a part of the PSP mission under
contract NNN06AA01C.  The \ISOIS data and visualization tools are
available to the community at:
https://spacephysics.princeton.edu/missions-instruments/isois; data
are also available via the NASA Space Physics Data Facility
(https://spdf.gsfc.nasa.gov/). Parker Solar Probe was designed, built,
and is now operated by the Johns Hopkins Applied Physics Laboratory as
part of NASA's Living with a Star (LWS) program (contract
NNN06AA01C). Support from the LWS management and technical team has
played a critical role in the success of the Parker Solar Probe
mission. We thank and acknowledge Dr. X.P. Zhao for providing the CSSS
model.

\bibliographystyle{apj}

\appendix

\section{Electron Observations}

Electron, Proton and Alpha Monitor (EPAM) observations
\cite[]{Gold:1998} on the ACE spacecraft at the Lagrangian 1 (L1)
point provide context for the observations made by PSP from April 18
through April 24, 2019. We use 20 min-average and fine-time resolution
measurements of the angular distribution of energetic electrons in the
energy range 45-290 keV detected by the sunward-looking telescope
LEFS60 of EPAM.  Inspection showed that the intensity profile of the
LEFS60 response tracks with the intensity profile of the magnetically
deflected electrons. Thus, the LEFS60 response is primarily due to
electrons.  The LEFS60 telescope has a geometrical factor equal to
$\sim$0.397 cm$^2$ sr. The number 60 (in 'LEFS60') denotes the angle
that the collimator centerline of the telescope makes with the
spacecraft spin axis.

The 20-min averaged measurements of electrons (DE) measured by the B
detector of the CA60 telescope are reported in the energy ranges
DE1 (38-53 keV), DE2 (53-103 keV), DE3 (103-175 keV) and DE4 (175-315
keV). The CA60 has a geometrical factor of 0.103 cm$^2$ sr. We also
present 20-min averaged measurements of energetic ion intensities from
ACE in the energy range 1.1-4.9 MeV as detected by the LEMS120
telescope.

\begin{figure}
\includegraphics[width=0.95\columnwidth]{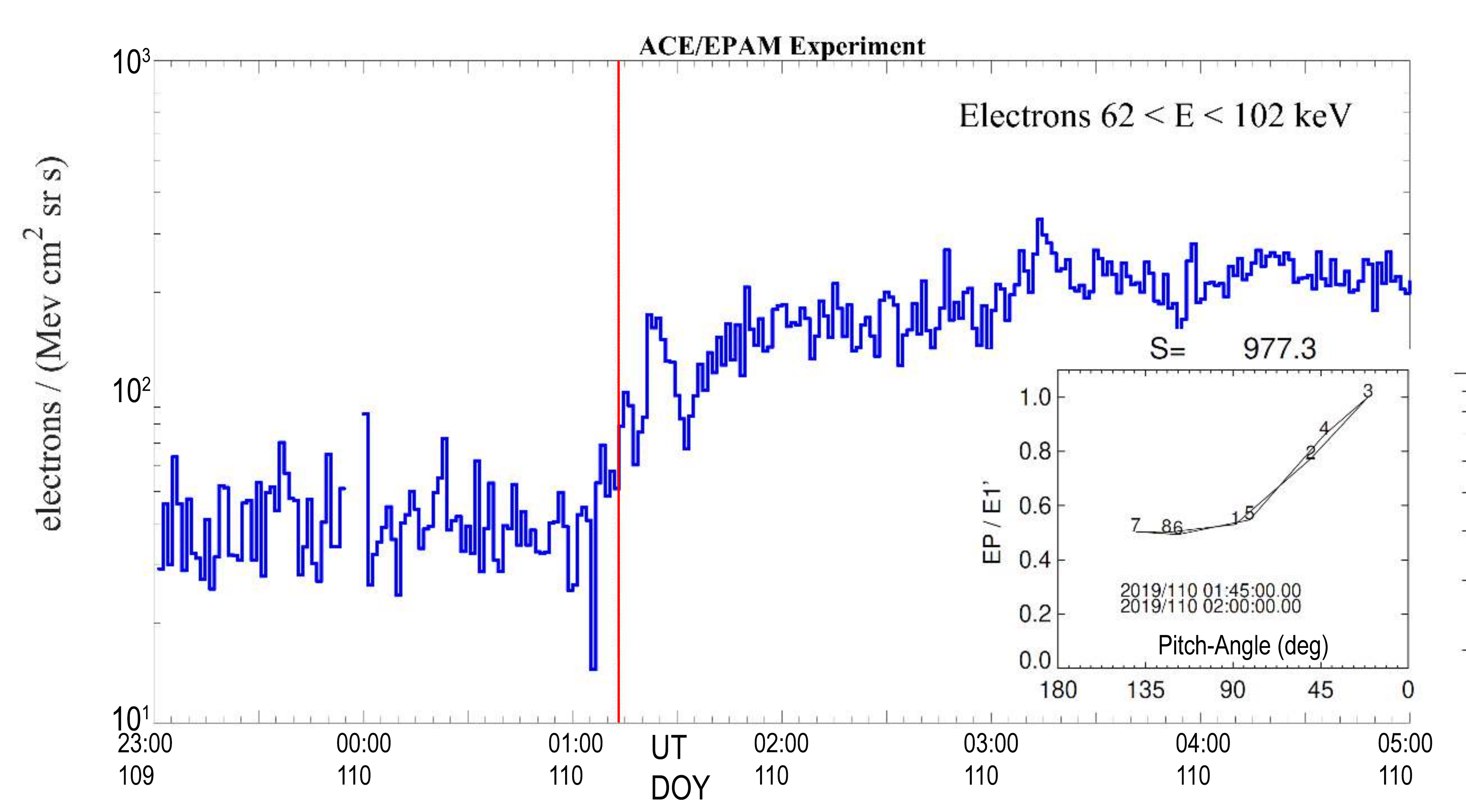} 
\caption{Observations of energetic electrons during the onset (red
  vertical line) of the April 20, 2019 event. The inset shows the
  maximum normalized intensity as a function of pitch-angle for the
  62-102 keV electrons streaming away from the Sun measured in one of
  the sectors of the LEFS60 sunward-looking telescope.  This is a
  representative snapshot of the highly anisotropic electron
  pitch-angle distributions observed during this event.  }
\label{fig:fluxElectronsApril20}
\end{figure}

Figure \ref{fig:fluxElectronsApril20} shows 80-sec averages of the
maximum intensity of the E'2 channel for the April 20, 2019 event.
These data correspond to 62-102 keV electrons measured in one of the
eight sunward-looking sectors of the LEFS60 telescope. In this case,
the E'2 channel is chosen for the determination of a clear onset. Even
if there were a residual straggling effect in this channel, this
instrumental effect leads to an underestimate of the delay for the
onset of this channel. Therefore, the onset in this channel gives us
(at the worst) a lower bound on the actual onset at those energies
\cite[]{Haggerty:2002}.  Since the energy spectrum of this particular
electron event is steep, the straggling effect of higher energy
channels on this channel is not expected to be significant.

Estimated instrumental background values have been subtracted from the
electron intensities.  The derived onset time of the electron event,
based on the 2$\sigma$ data-driven onset time determination method
\cite[see][for more details for this onset time determination
  method]{Malandraki:2012} is marked by the red vertical line at 01:13
UTC on April 20, 2019. Taking the FWHM of the electron Pitch-Angle
Distributions (PADs) at 01:45 UTC, an effective pitch angle of
45$^\circ$ is obtained. The transit time for a 45$^\circ$ pitch-angle
along a nominal 1.2 AU long Parker spiral, for the mean energy of the
E'2 channel, is $\sim$29 minutes; therefore the anticipated electron
release time at the Sun is found to be 00:44 UTC on April 20. This
release time is roughly 40 minutes prior to the corresponding CME
release time listed in Table \ref{tab:table}, implying that the
electron release likely occurred near the CME initiation period.

The briefest of the three events began on April 20, around 09:15 UTC
during the decay phase of the event that occurred earlier (00:44 UTC)
that day. This electron event did not have a clear association with an
ion event, but the onset time lines up with the second Type III
event that occurred on April 20. 

\begin{figure}
\includegraphics[width=0.95\columnwidth]{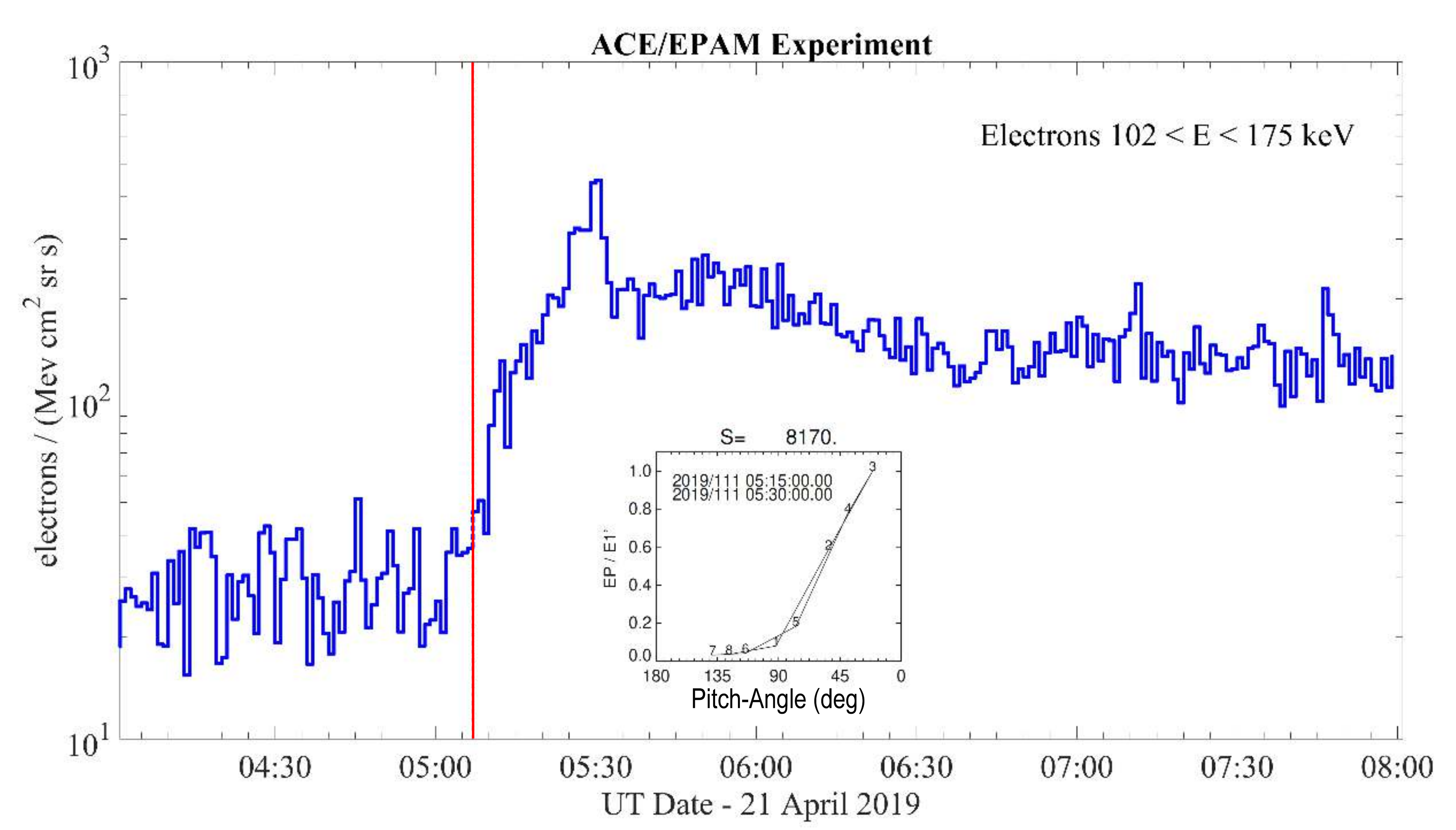} 
\caption{Observations of energetic electrons during the onset of the
  April 21, 2019 event. The maximum intensity of the 102-175 keV electrons
  streaming away from the Sun measured in one of the sectors of
  the sunward-looking telescope LEFS60 for each 1-min interval is
  shown. The inset presents 15-min averaged PADs of 45-62 keV
  electrons at 05:15 UT April 21, 2019, which exhibit a strong
  anisotropy directed parallel to the IMF. Normalized differential
  intensity is plotted versus pitch-angle. The red vertical
  line indicates the time of the determined onset of the electron
  event.}
\label{fig:fluxElectronsApril21}
\end{figure}

In Figure \ref{fig:fluxElectronsApril21}, we show 1-min averages of
the maximum intensity of the E'3 channel during the April 21, 2019
event. These data corresponds to 102-175 keV energetic electrons that
stream away from the Sun as measured in one of the eight
sunward-looking sectors of the LEFS60 telescope. In this event, the
two lower energy electron channels (E'1 and E'2) have an enhanced
ambient flux due to a prior weak electron event, which masks the event
onset fluxes and makes it difficult to accurately determine onset
times in these channels. Furthermore, these two channels can be
strongly affected by straggling of the higher energy electrons
(depending on the steepness of the spectrum), whereas this effect is
negligible in the highest two channels (E'3 and E'4), as previously
highlighted by \cite{Haggerty:2002} and \cite{Haggerty:2003}. Since
for this electron event the E'4 electron channel enhancement was
rather weak, we have utilized the E'3 channel measurements as the
highest energy channel (i.e. highest velocity electrons) where an
onset time can be reliably determined.

The red vertical line marks the time of the onset of the electron
event, at 05:07 UTC, which was determined as the time that the
intensity exceeded the background level by 2$\sigma$
\cite[]{Malandraki:2012}. The inset of Figure
\ref{fig:fluxElectronsApril21} shows 15-min averaged PADs for the
45-62 keV electrons at 05:15 UTC on April, 21, 2019. On that day the
radial component of the Interplanetary Magnetic Field (IMF) was
pointing away from the Sun. The electron population comprises an
electron beam that exhibits a strong anisotropy directed parallel to
the IMF and is therefore propagating away from the Sun. These peaked
electron PADs argue for nearly scatter-free propagation of these
particles during their outward transit from the corona to the ACE
spacecraft at L1. Taking the FWHM of the electron PADs at 05:15 UT on
April 21, 2019 we obtain a value of 35$^\circ$ as the effective pitch
angle of this electron population. The transit time of an electron
with a 35$^\circ$ pitch-angle, along a nominal 1.2 au Parker
spiral length, and with a mean energy of the E'3 channel is $\sim$20
minutes. Therefore, the deduced electron injection time at the Sun is
found to be at 04:47 UTC on April 21, 2019, which is 13 minutes prior
to the CME release time listed in Table \ref{tab:table}.

\section{Current Sheet Source Surface Model}
\label{sec:CSSS}

We used the Current Sheet Source Surface (CSSS) model
\citep{zha95,pod14,pod16,jac19} of the corona in magnetostatic
equilibrium \citep{bog86}. The analytical solutions \cite[]{bog86}
incorporate volume and sheet currents effectively \cite[see][and the
  references therein]{zha95} by dividing the corona into three regions
separated by two concentric spherical surfaces; the inner surface, the
cusp surface (associated with the cusps of helmet streamers) placed at
around 2.5 $R_s$, and the outer source surface \cite[Figure~1
  in][]{zha95} placed at $\sim 15 R_s$.  The magnetic field lines are
open at the cusp surface though still non--radial until the source
surface where the solar wind moves radially out into the heliosphere.

The CSSS model extrapolates the observed photospheric magnetic field
to obtain the coronal magnetic field. The model takes the synoptic map
of the photospheric magnetic field as input and employs a spherical
harmonic expansion to compute the coronal magnetic field for the
inner, middle and outer coronal regions separated by the cusp and
source surfaces. For the mapping presented in Figure
\ref{fig:backmapping}, we used the synoptic map constructed using the
high--resolution, high--cadence magnetograms taken by the Helioseismic
and Magnetic Imager (HMI) telescope on board the Solar Dynamics
Observatory (SDO) for Carrington Rotation CR2216 (April~8--May~5,
2019).

\section{Enlil Modeling of the April 21, 2019 CME}
\label{sec:enlilApp}

Enlil is a time-dependent 3D MHD model of the heliosphere
\cite[]{Odstrcil:2003}. It solves equations for plasma mass, momentum
and energy density, and magnetic field, using a
Flux-Corrected-Transport (FCT) algorithm.  The Enlil cone model
forecasts CME propagation from the Enlil inner boundary (at 21.5 R$_s$)
to 2 au.  The ambient solar wind is based on the  Wang-Sheeley-Arge
(WSA) model \cite[]{Arge:2000}. In the Enlil cone model, the CME
propagates out close to the Sun with constant angular and radial
velocity. The Enlil model takes the following input parameters at its inner
boundary
\begin{itemize}
\item \textbf{Start time at 21.5 R$_s$:}  April 20, 2019 at 10:33 UTC.
\item \textbf{Direction:} HEEQ longitude 90$^\circ$ and  latitude 2$^\circ$.
\item \textbf{Half Angular Width:} 30$^\circ$ half of the full angular
  width of the cone.
\item \textbf{Speed:} 387 km s$^{-1}$ radial velocity (km/s) at the
  Enlil inner boundary
\end{itemize}

\begin{figure}
\includegraphics[width=0.95\columnwidth]{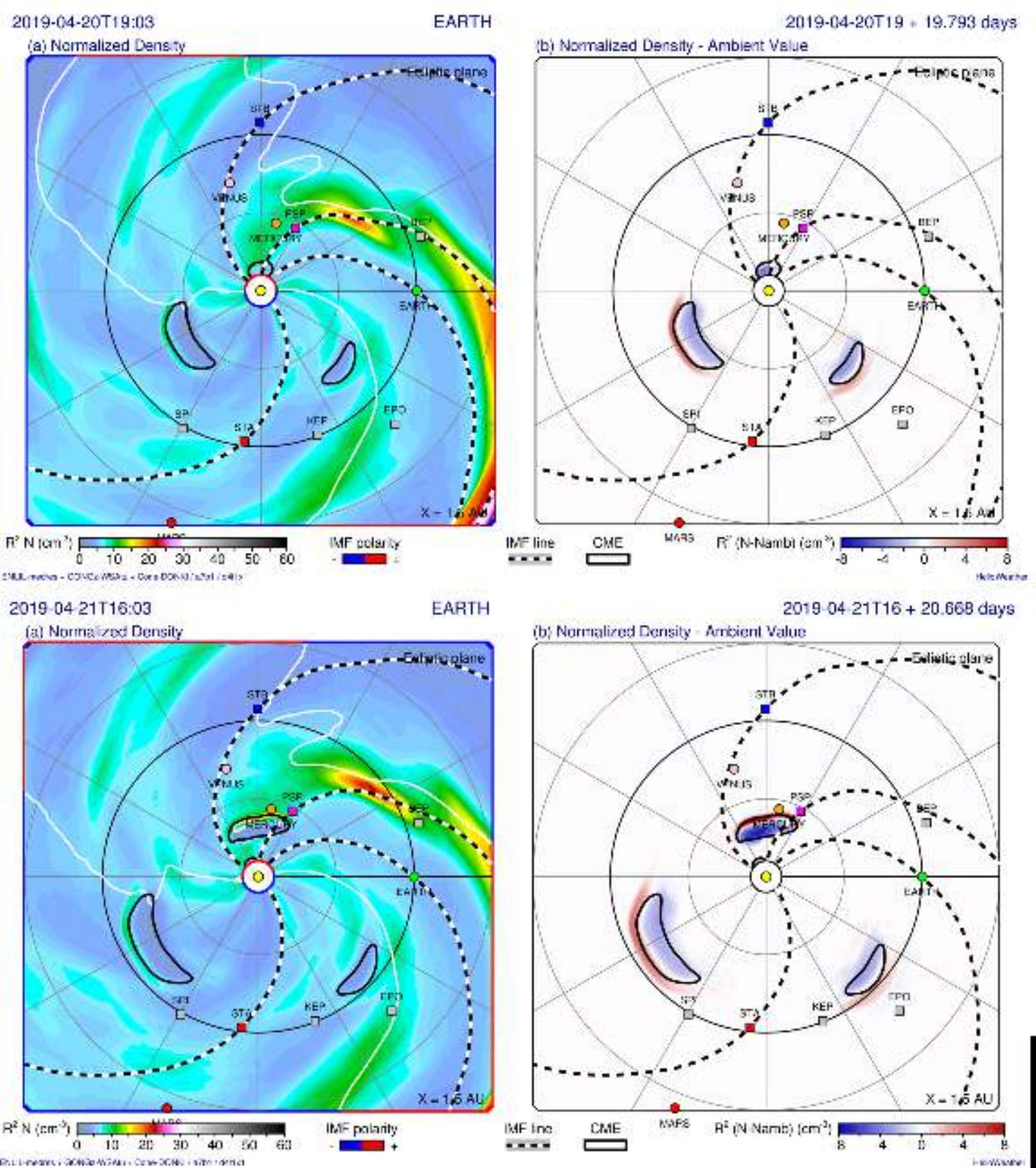} 
\caption{Snapshots showing the Enlil model of the April 20, 2019 CME
  propagating out to PSP.  Left panels show modeled densities and
  right panels show the difference between the modeled density with
  the CME and the ambient density in the background solar wind.  }
\label{fig:enlilc1}
\end{figure}

\begin{figure}
\includegraphics[width=0.95\columnwidth]{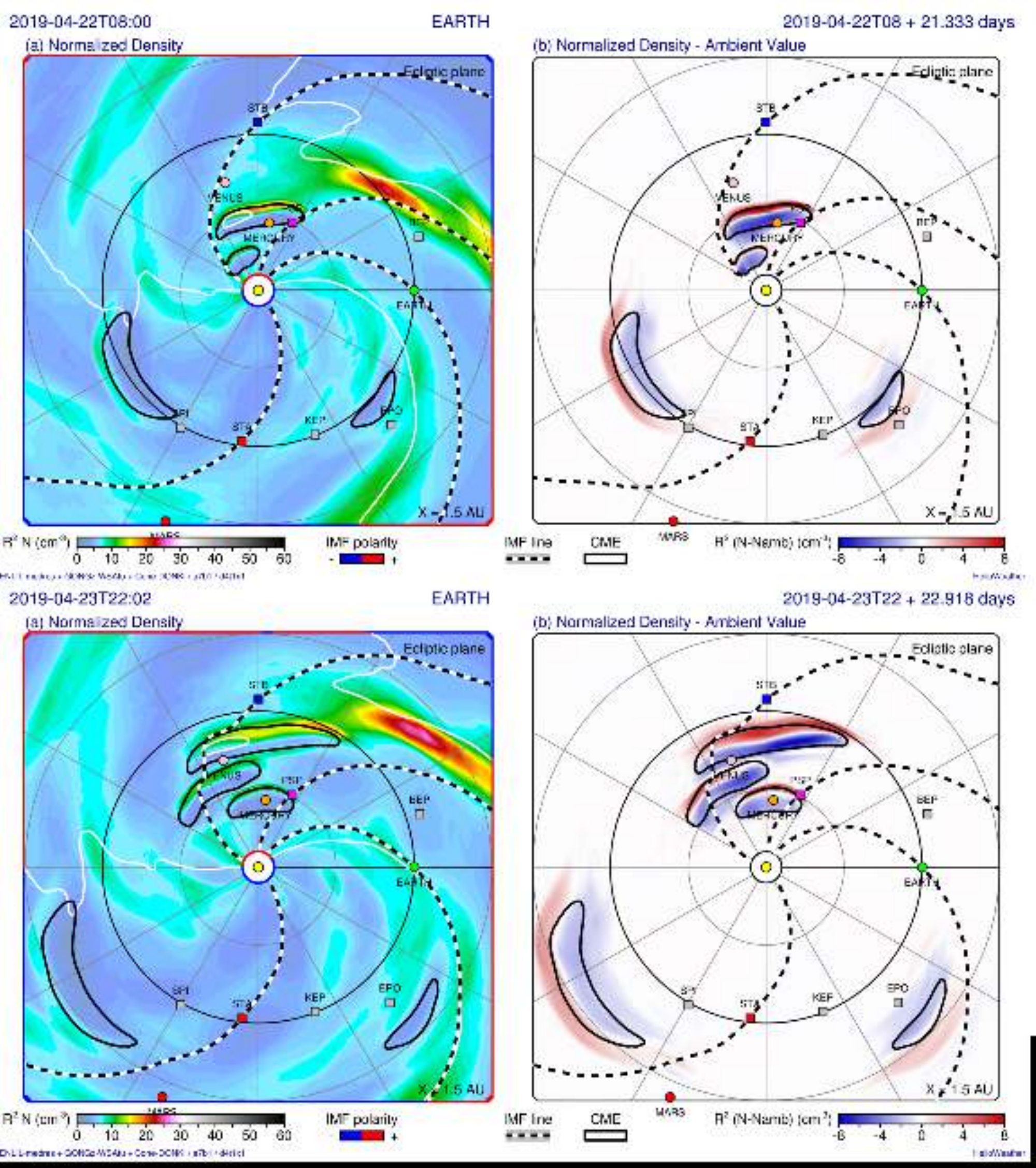} 
\caption{Enlil model snapshots similar to Figure \ref{fig:enlilc1}
  showing the stages of the CME's propagation after April 21 out to 1
  au.}
\label{fig:enlilc2}
\end{figure}

\begin{figure}
\includegraphics[width=0.9\columnwidth]{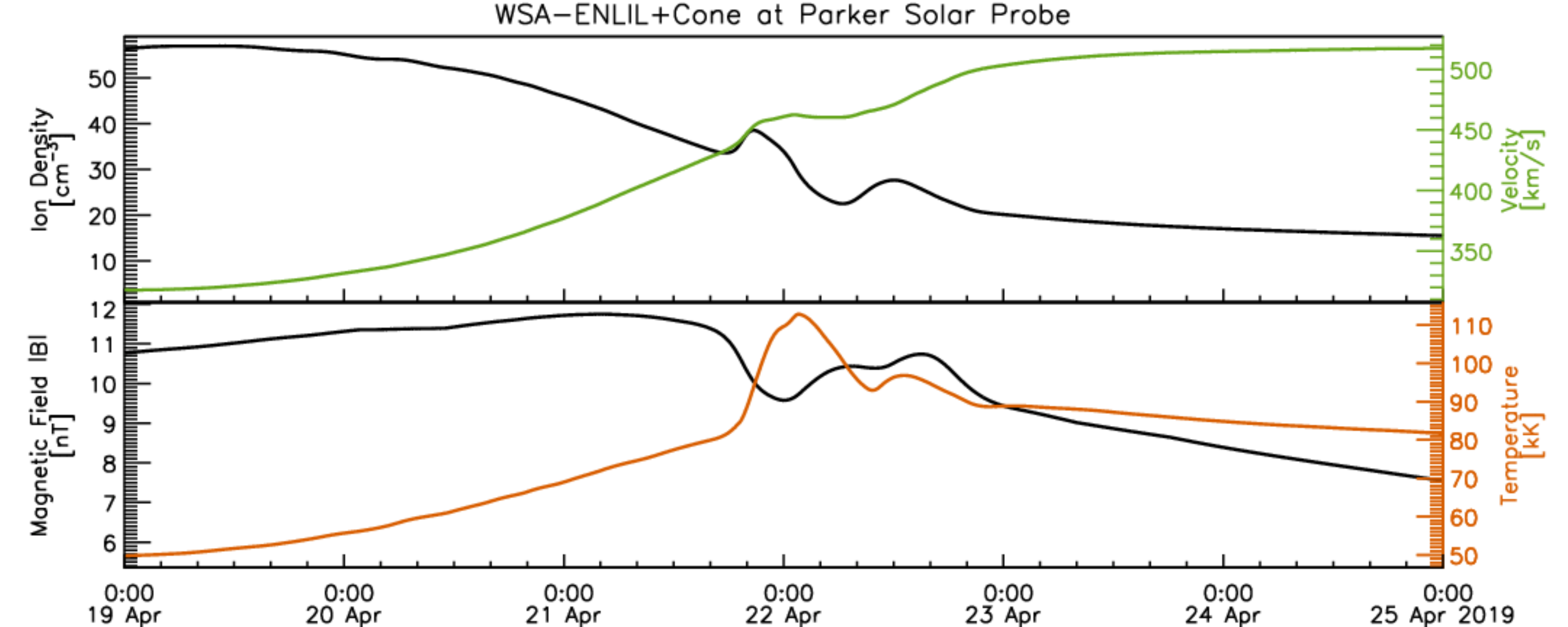} 
\caption{Timeline of the Enlil model results at the location of the
  PSP spacecraft: (top panel) ion density (black), solar wind and CME
  velocity (green); (bottom panel) magnetic field strength (black),
  and plasma temperature (orange).  }
\label{fig:enlilc3}
\end{figure}

Figures \ref{fig:enlilc1} and \ref{fig:enlilc2} show the evolution of
the CME released on April 20. The structure propagates to and beyond
PSP, driving a large compression in front of the structure. These
images are taken from a movie available as supplementary
material. Figure \ref{fig:enlilc3} shows the timeline of the modeled
plasma at the location of the PSP spacecraft.

\section{Compressive Enhancement of Energetic Particles}
\label{sec:compress}

\begin{figure}
\includegraphics[width=0.7\columnwidth]{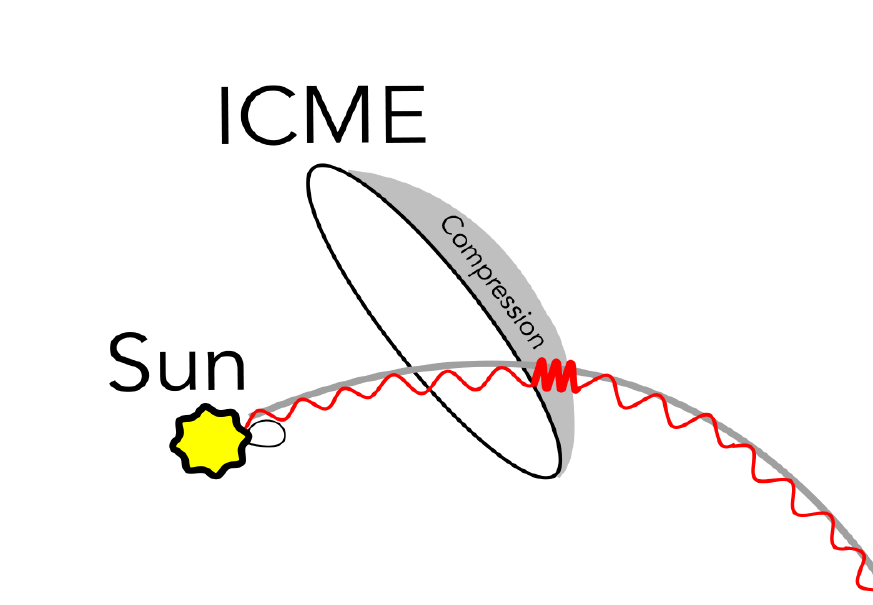}
\caption{Illustration of a CME-driven compression region including
  energetic particles that diffuse away from the Sun after a flare or
  particle acceleration from the low corona.  }
\label{fig:icmeApp}
\end{figure}

In this appendix, we describe a scenario in which energetic particles
are compressed by the plasma as they move into a compression region,
as illustrated in Figure \ref{fig:icmeApp}. Our treatment departs from
the conventional solution for DSA. We
begin by describing how diffusive ramps are treated within DSA theory
and show how compressive enhancements differ.

For simplicity, we take the compression formed simply from faster
solar wind plasma with speed $u_f$ ramming into a slower plasma with
speed $u_s$. The compression ratio is $r_c = u_f/u_s$ and the local
width of compression region with a speed gradient is taken to be
$\delta x$.  We assume that the magnetic field and the solar wind
direction are aligned in the $x$-direction.

The evolution of the isotropic part of the distribution function is
typically described using the Parker transport equation
\cite[]{Parker:1965}:
\begin{eqnarray}
\frac{\partial f_0}{\partial t} + \mathbf{u} \cdot \nabla f_0 - \nabla \cdot \left( \mathbf{K} \cdot \nabla f_0 \right) - \frac{\nabla \cdot \mathbf{u}}{3} p \frac{\partial f_0}{\partial p} = 0
\end{eqnarray}
where $\mathbf{u}$ is the solar wind velocity, and $\mathbf{K}$ is the
diffusion tensor. In DSA theory, a discontinuity between fast and slow
wind creates the conditions for rapid particle acceleration. Upstream
from the fast-slow wind interface, there is a balance of diffusive
streaming against the convected fast solar wind, 
\begin{eqnarray}
u_f f_0 - \kappa  \frac{\partial f_0}{\partial x} = 0.
\end{eqnarray}
The convective-diffusion solution follows, 
\begin{eqnarray}
f_{u}(x) = f_{s} \exp( x u_f/\kappa)
\label{eq:diff1}
\end{eqnarray}
where $f_s$ is isotropic part of the distribution at the stream
interface, $x=0$, and $f_u$ is the upstream solution.  The upstream
solution in Equation (\ref{eq:diff1}) results from outward convection
of energetic particles away from the Sun and diffusive streaming back
toward the Sun. It is important to note, however, that we have assumed the
source is a delta function at $x=0$. If there is a seed population,
then the correct upstream solution involves both the shock-accelerated
population that streams against the flow and the pre-existing seed
population that convects with the flow.

We consider a different scenario in which the gradient in solar wind
speed is not a discontinuity. We take the distribution to remain
approximately isotropic as it is convected into the speed gradient
such that the rate of convection exceeds the rate of diffusion. In
this departure from DSA theory, we take
\begin{eqnarray}
u \frac{\partial f_0}{\partial x} > \frac{\partial}{\partial x}\left(\kappa \frac{\partial f_0}{\partial x} \right).
\end{eqnarray}
This requires that $u > \kappa/\delta x$. Equivalently, this places a requirement on 
the scattering mean free path, 
\begin{eqnarray}
\lambda <  3 u \delta x / v.  
\label{eq:lam2}
\end{eqnarray}
Taking $u = 400$ km s$^{-1}$, a width of $\delta x = 0.1$ au, and a 1
MeV proton, this would require a scattering mean free path, $\lambda <
2$ R$_s$ (or 0.009 au) within the compression.

To help conceptualize the limit on the mean free path, we consider the
rate of particle acceleration if the compression were instead a
discontinuity.  The DSA acceleration time to a given momentum $p$ is
\cite[]{Forman:1983, Drury:1983, Jokipii:1982, Jokipii:1987, Schwadron:2015SEP}:
\begin{eqnarray}
\tau_p \approx \frac{3 \delta x_\mathrm{dsa}}{\Delta u} 
\label{eq:tdsa}
\end{eqnarray} 
where $\Delta u = u_f - u_s$, the width of the DSA acceleration region is 
\begin{eqnarray}
\delta x_\mathrm{dsa} = \frac{\kappa_f}{u_f} + \frac{\kappa_s}{u_s} ,
\end{eqnarray}
and $\kappa_s$ and $\kappa_f$ are the diffusion coefficients on the
slow and fast wind side of the stream discontinuity. For simplicity,
we take $\kappa_f > \kappa_s \approx 0$ due to the presence of downstream
turbulence.  Using the
previously stated limit for $\lambda$, Equation (\ref{eq:lam2}), we find
that
\begin{eqnarray}
\delta x_\mathrm{dsa} < \delta x.
\label{eq:dxdsa}
\end{eqnarray}
Not surprisingly, this implies that the region of diffusive
acceleration must be smaller than the width of 
compressive gradient. Conversely, if the width of the speed 
gradient exceeds the width of DSA acceleration region, then 
compressive acceleration will dominate. 

We find the DSA acceleration time by substituting Equation
(\ref{eq:dxdsa}) into Equation (\ref{eq:tdsa}),
\begin{eqnarray}
\tau_p < 3 \frac{\delta x}{\Delta u}. 
\label{eq:tp}
\end{eqnarray}
The quantity $\delta x/\Delta u$ represents the convection time
through the speed gradient. Therefore the condition in Equation
(\ref{eq:tp}) implies that if the convection time exceeds  the DSA
acceleration time, then compressive acceleration will dominate. 

Given the condition on the scattering mean free path in Equation
(\ref{eq:lam2}), we find the following approximation for transport 
into the compression,
\begin{eqnarray}
 u \frac{\partial f_0}{\partial x} - \frac{1}{3}\frac{\partial u}{\partial x} p \frac{\partial f_0}{\partial p} = 0. 
\end{eqnarray}
This is solved for the compressed isotropic portion of the
distribution function, $f_c$, as a function of the distribution
function $\tilde{f}_0$ convected into the compression,
\begin{eqnarray}
f_c = r_c^{\gamma/3} \tilde{f}_0.
\label{eq:sol}
\end{eqnarray}
Here $\gamma$ is the power-law index of the energetic particle
distribution in the faster wind behind the  compression, 
$\tilde{f}_0 \propto p^{-\gamma}$.

\end{document}